\documentclass[aps,onecolumn,preprintnumbers,superscriptaddress,nofootinbib,preprint]{revtex4}
\usepackage{epsfig,float,amssymb,latexsym,amsmath,amsthm,fancyhdr}
\usepackage{graphics,psfrag,longtable}
\usepackage{slashed}
\usepackage[T1]{fontenc}
\usepackage{xcolor}

\usepackage{hyperref}

\begin{document}

 \author{J. M. Alarc\'on}
  \email{jmanuel.alarcon@uah.es}
\affiliation{University of Alcal\'a,
Nuclear and Particle Physics Group, Department of Physics and Mathematics, Alcal\'a de Henares (Madrid), Spain}

\title{Brief history of the pion-nucleon sigma term}

\begin{abstract}
%\textbf{Abstract.}
The pion-nucleon sigma term is a quantity involved in many important aspects of particle and nuclear physics. 
In this review I show its origin and how it is connected to important questions as the origin of mass of the ordinary matter, studies of dark matter detection and nucleosynthesis. 
I mention the most common methods used to obtain the sigma term, and comment on the extracted values. 
As it is shown, the accepted value of the sigma term has been moving from relatively low ($\sim 40$~MeV) to larger ones ($\sim 60$~MeV) until today, where this controversy still persist. 
\end{abstract}

%\keywords{???}

\maketitle
%
%\tableofcontents

\section{Introduction} 
\label{Sec:Introduction}

Understanding the strong interactions has been one of the main goals in theoretical physics in the last decades. 
Experimental information on the strongly interacting particles allowed to infer some important features of a possible fundamental theory, even before Quantum Chromodynamics (QCD) was formulated.
One of this features is the approximated chiral symmetry of this interaction, that permits to obtain interesting results concerning the strongly interacting particles at low energy (low-energy theorems) \cite{Adler,Weinberg,GellMann:1968rz}.
On the other hand, experimental tests of the low-energy theorems was a possible way to obtain information of a fundamental theory.

The $\pi N$ scattering was considered for long time the ideal process to gain insight into the strong interactions since, experimentally, was in much better situation than other hadronic processes. 
For this process, one can derive a low-energy theorem that involves the symmetry breaking part of a fundamental theory for strongly interacting particles as follows:
Let's consider the process

\begin{align}
 \pi ^a (q) N(p) \to  \pi^b (q') N(p')
\end{align} 

where $a\ (b)$ are the isospin indices of the incoming (outcoming) pion with momentum $q\ (q')$, and $p\ (p')$ the momentum of the incoming (outcoming) nucleon.

We can write the amplitude of this process for pions with masses $q^2$ and $q'^2$ in the following way \cite{Reya:1974gk}

\begin{align}\label{Eq:T2}
 T^{ba}(\nu, t, q^2, q'^2) = i \frac{(q^2 - M_\pi^2)(q'^2 - M_\pi^2)}{M_\pi^4 f_\pi^2} \int d^4x \langle N(p')| T\left( \partial_\mu A^{b \mu}(x) \partial_\nu A^{a\nu}(0)\right)|N(p) \rangle e^{i q' x}, 
\end{align} 

where $\nu = \frac{s - u}{4 m_N}$, being $m_N$ the nucleon mass, $A^\mu$ the axial current, $f_\pi$ the weak decay constant of the pion and $M_\pi$ the pion mass.
Pulling the derivatives out of the time-orderd product, we arrive to the so-called generalized Ward-Takahashi identity for this particular case, which has the form

{\small
\begin{align}\label{Eq:WT}
&\int d^4x \langle N(p')| T\left( \partial_\mu A^{b \mu}(x) \partial_\nu A^{a\nu}(0)\right)|N(p) \rangle e^{i q' x} = \int d^4x \langle N(p')| \left\{ q'_\mu q_\nu T\left(  A^{b \mu}(x)  A^{a\nu}(0)\right) \right. \nonumber \\
& \left.+ i q'_\mu \delta(x_0) [A^{b \mu}(x) , A^{a\nu}(0)]- \delta(x_0) [A^{b 0}(x) , \partial_\nu A^{a\nu}(0)] \right\} |N(p) \rangle e^{i q' x}. 
\end{align}
}

Note that the last term is not determined by current algebra, and can be isolated by taking the so-called ``soft-meson limit" $q'_\mu, q_\nu \to 0$. 
So, using Eq.~\eqref{Eq:WT} in Eq.~\eqref{Eq:T2} and taking that limit, one has

\begin{align}\label{Eq:Def-sigma-commutator1}
 T^{ba}(0, 0, 0, 0) &= - \frac{1}{f_\pi^2}\int d^4x \langle N(p')| \delta(x_0)  i [A^{b 0}(x) , \partial_\nu A^{a\nu}(0)] |N(p) \rangle \nonumber \\
                            &=  - \frac{1}{f_\pi^2} \langle N(p')| [Q_A^{b}(0) , [Q_A^{a}(0), \mathcal{H}_{\chi SB} ]]|N(p) \rangle
\end{align} 

where the $Q_A^i$ are the axial charges and $ \mathcal{H}_{\chi SB}$ the Hamiltonian density that breaks the chiral symmetry of the strong interactions. 
The commutator in the last matrix element of Eq.~\eqref{Eq:Def-sigma-commutator1} is the so-called {\it sigma commutator}, and inherits this name from the $\sigma$-model, where it reduces to the $\sigma$ field. 
Using the Jacobi identity, one sees that the sigma commutator is symmetric in the isospin indices and, therefore, contributes only to the isoscalar part of the scattering amplitude, $T^+$.
QCD provides an explicit form for $\mathcal{H}_{\chi SB}$ (the mass term), what allows to relate the sigma commutator in the $SU(2)$ case to the matrix element

\begin{align}
\hspace{4cm} \sigma_{\pi N} =\frac{\hat{m}}{2 m_N} \langle N(p)| \bar{u}u + \bar{d}d  |N(p) \rangle \hspace{0.5cm} \left(\hat{m} = \frac{m_u+ m_d}{2}\right),
\end{align} 

usually called the {\it pion-nucleon sigma term}, or simply, the sigma term\footnote{In the previous equation the $2 m_N$ in the denominator appears because I used the normalization $\langle N(p')|  N(p)\rangle = (2\pi)^3 2E\, \delta({\vec{p}\,' - \vec{p}})$, what will be customary for the rest of the article. Notice also that $\hat{m}$ arises because we are working in the $SU(2)$ symmetric limit.}. 

\begin{figure}[H]
 \begin{center}
 \includegraphics[width=.9\textwidth]{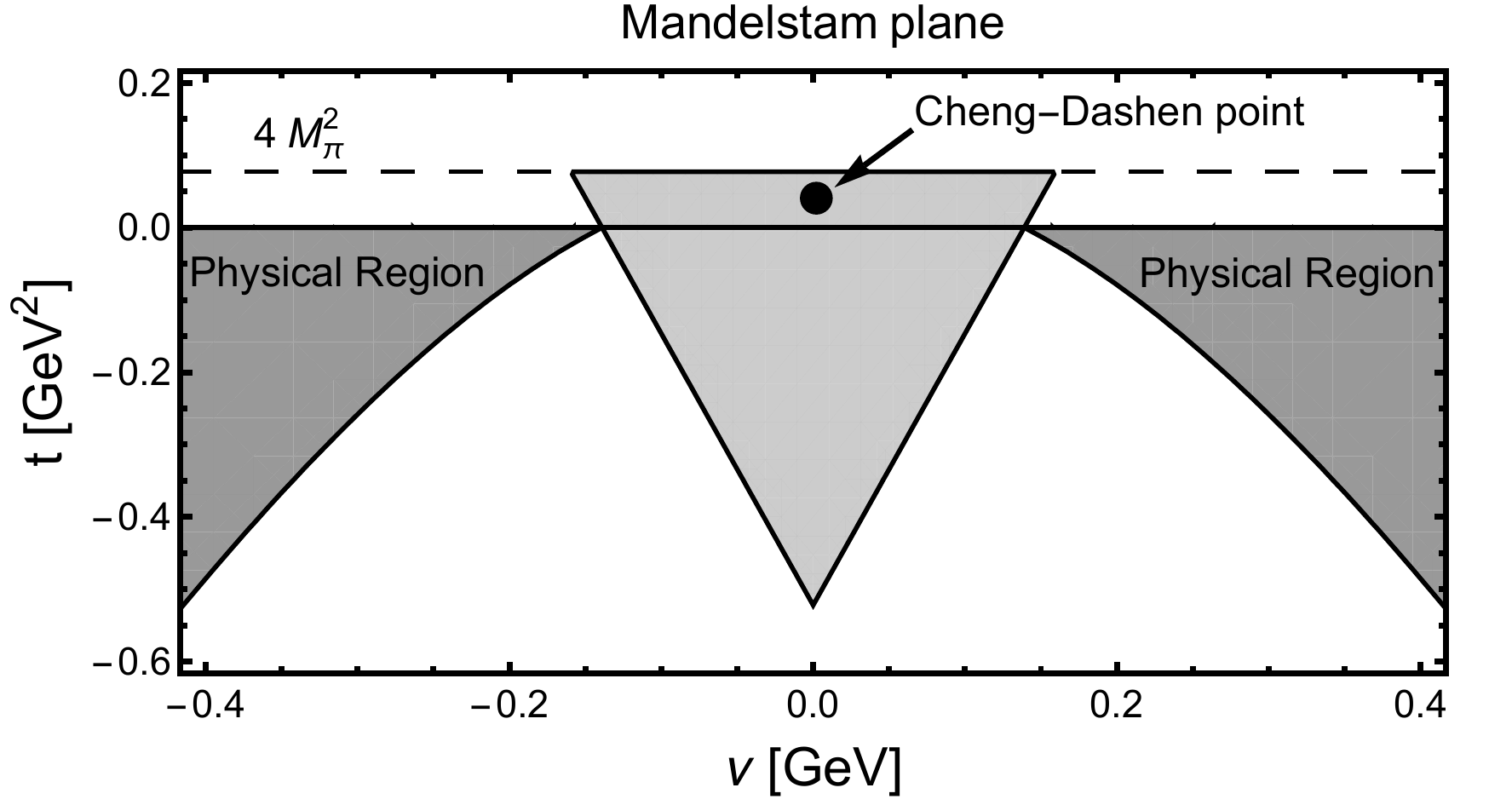} 
 \caption{Kinematic regions for $\pi N$ scattering. The light gray area (Mandelstam triangle) correspond to the region $s < (m_N+M_\pi)^2$, $u < (m_N+M_\pi)^2$ and $t < 4M_\pi^2$, where the Born-subtracted amplitudes are analytic and real. The gray area is the physical region. The dashed line corresponds to $t = 4M_\pi^2$.}
 \label{Fig:Mandelstam_plane}
 \end{center}
\end{figure}

The interest in this low-energy theorem in the early times of current algebra was that allowed to test the models for chiral symmetry breaking through the knowledge of  $T^{ba}(0, 0, 0, 0)$. 
Nowadays its interest is more related to the form of the QCD operator that is evaluated between nucleon states, what makes it a very relevant quantity in many areas of particle and nuclear physics, as I will show in Sec.~\ref{Sec2}. 
This amplitude, however, is evaluated at $\nu = 0$ and $t = 0$ for {\it unphysical} meson masses. 
One needs therefore a way to connect the experimental data (with physical pion masses) to the unphysical scattering amplitude $T^{ba}(0, 0, 0, 0)$. 
One possible strategy is to use dispersion relations in the mass variable to extrapolate the low-energy theorem for unphysical pions with zero momentum to physical ones, as was done in Ref.~\cite{Fubini&Furlan}.
However, there is a simpler way to relate the amplitude with physical pions to its soft-pion limit. 
As was shown in Ref.~\cite{Cheng:1970mx}, one can expand the amplitude in $q^2$ and $q'^2$ to connect the physical amplitude to its soft limit. 
By invoking the Adler consistency conditions and evaluating the amplitude at $(\nu = 0, t = 2M_\pi^2)$ it is possible to cancel the leading correction in $q^2$ and $q'^2$, and relate the physical amplitude to $T^{ba}(0, 0, 0, 0)$ plus a remainder. 
This kinematical point is called the Cheng-Dashen point, and lies in the unphysical (subthreshold) region of $\pi N$ scattering (see Fig.~\ref{Fig:Mandelstam_plane}).
In the original work of Cheng and Dashen, it was assumed that the remainder was of order $M_\pi^4$. 
Later, Brown, Pardee and Peccei showed that there were non-analytic contributions that made this relation valid only up to order $M_\pi^2$ \cite{Brown:1971pn}. 
They also presented a modified form of the Cheng-Dashen theorem that was valid up to $\mathcal{O}(M_\pi^4)$, although did not involved explicitly $\sigma_{\pi N}$.
To be more specific, if one writes the $\pi N$ scattering amplitude in terms of the four Lorentz invariant amplitudes $A^\pm$, $B^\pm$

\begin{align}\label{Eq:T-decomposition}
 &T^{ba} = \delta^{ba} T^+ + \frac{1}{2}\left[ \tau^b, \tau^a\right] T^- \\
 &T^\pm  = \bar{u}' \left[  A^\pm + \frac{1}{2} (\slashed{q}' + \slashed{q} )B^\pm \right] u =  \bar{u}' \left[  D^\pm - \frac{1}{4m_N} [\slashed{q}' + \slashed{q} ]B^\pm \right] u
\end{align}

where $ T^\pm$ are the isoscalar and isovector parts of the amplitude and $D^\pm \equiv A^\pm + \nu B^\pm$, the original result by Cheng and Dashen states that $\Sigma\equiv f_\pi^2  \bar{D}^+(\nu = 0, t = 2 M_\pi^2)$ and $\sigma_{\pi N}$ are equal up to order $M_\pi^2$, being $\bar{D}^+$ the Born-subtracted part of the amplitude.

In Ref.~\cite{Brown:1971pn} this theorem was recasted as follows

\begin{align}\label{Eq:Pagels&Pardee}
 f_\pi^2  \bar{D}^+(\nu = 0, t = 2 M_\pi^2) = \sigma(t = 2 M_\pi^2) + \Delta_R,
\end{align}

where 

\begin{align}
\sigma(t) = \frac{1}{2 m_N} \langle N(p')| \hat{m}(\bar{u}u + \bar{d}d)  |N(p) \rangle.
\end{align}

This form has the advantage that (i) $\Delta_R$ is of order $M_\pi^4$ and small ($\Delta_R \simeq 2$~MeV \cite{Bernard:1996nu}), and (ii) 
the scalar form factor of the nucleon at $t = 2M_\pi^2$, $\sigma(t = 2M_\pi^2)$, can be related to the sigma term though dispersion relations or other methods, since $\sigma(t=0) = \sigma_{\pi N}$.
The latter was calculated in Ref.~\cite{Pagels:1972kh}, with a result of $\Delta_\sigma \equiv \sigma(t = 2M_\pi^2) - \sigma_{\pi N} \approx 14$~MeV.
As we will see later, $\Delta_\sigma$ has been updated a couple of times, and became an important part of the  controversy on the value of $\sigma_{\pi N}$.

As we have seen, the low energy theorem derived from symmetry arguments allows to connect experimental data to the symmetry breaking part of the Hamiltonian describing the strong interactions. 
If that Hamiltonian is taken from QCD, then one has access to the scalar matrix element of QCD between nucleon states, which is extremely important in many areas of particle an nuclear physics, as I will show in the following section.

\newpage
\section{The sigma term in particle and nuclear physics}
\label{Sec2}

Most of the interest in the sigma term has to do with its relation to important questions in particle and nuclear physics. 
In this section I will mention some of them.

\subsection{Origin of the mass of the ordinary matter}

In the last years there has been an increasing effort to understand the basic properties of hadrons in terms of QCD. 
A prominent example is explanation of the proton (or nucleon) mass in terms of QCD degrees of freedom. 
The usual strategy to analyze its decomposition is to consider that the energy-momentum tensor of QCD can be related to the mass of a hadron (in this case the nucleon) in the following way\footnote{Note that the $1/2m_N$ factor comes from the normalization of the nucleon states.}

\begin{align}\label{Eq:nucleon-mass-1}
 m_N = \frac{1}{2m_N}\langle N(p) | T^\mu_{\ \mu} | N(p) \rangle 
\end{align} 

where $T^\mu_{\ \mu}$ is the trace of the energy momentum tensor.
This trace has been worked out some time ago \cite{Crewther:1972kn,Chanowitz:1972vd,Collins:1976yq,Shifman:1978zn} and, using its renormalized form, the matrix element \eqref{Eq:nucleon-mass-1} becomes

\begin{align}\label{Eq:nucleon-mass-2}
 m_N &= \frac{1}{2m_N}\langle N(p) | \frac{\beta(\alpha_s)}{2\alpha_s} G^{a \mu \eta }G^{a }_{\ \mu\eta} + \sum_{q = u, d, s} (1+\gamma_{m_{q}}) m_q \bar{q}q  | N(p) \rangle   \nonumber \\
 & = \frac{1}{2m_N}\langle N(p) | \frac{\beta(\alpha_s)}{2\alpha_s} G^{a \mu \eta }G^{a}_{\ \mu \eta} + \sum_{q = u, d, s} \gamma_{m_q} m_q \bar{q} q   | N(p) \rangle \nonumber \\
 &+  \frac{1}{2m_N}\langle N(p) | \sum_{q = u, d, s} m_q\bar{q} q  | N(p) \rangle,
\end{align}

where $q$ and $G^{a \mu \nu }$ are the quark and gluon fields, $\alpha_s$ is the strong coupling, $\beta(\alpha_s)$ its beta function, $m_q$ is the mass of the quark $q$ and $\gamma_{m_q}$ the anomalous dimension of the mass operator.
In the $SU(2)$ limit, the las term in \eqref{Eq:nucleon-mass-2} is the sigma-term\footnote{The other matrix element, $\langle N(p) | m_s\bar{s}s   | N(p)\rangle$ can be estimated with the help of $\sigma_{\pi N}$, as is shown later.}. 
Notice that this term is by itself renormalization scheme and scale independent, and can be interpreted as the light quark mass contribution to the nucleon mass. 
In this sense, this contribution has its origin in the Brout-Englert-Higgs mechanism. 
Therefore, an accurate determination of $\sigma_{\pi N}$ is important to know how much of the ordinary matter mass is generated though this mechanism and how much is dynamically generated.

\subsection{Dark matter detection}

In the last years, the main interest for a precise determination of $\sigma_{\pi N}$ is related to its role in dark  matter searches. 
In particular, for scalar dark matter detection.
The reason is because, if one assumes dark matter of scalar nature, we can construct an effective Lagrangian that couples dark matter ($\chi$) to quarks ($q$) as follows,

\begin{align}
 \mathcal{L}_{\chi q} = C \frac{m_q}{\Lambda^3} \bar{\chi} \chi \bar{q}q
\end{align}

where $C$ is a Wilson coefficient that contains information of the energy scales higher than the cutoff $\Lambda$. 
Assuming that $\chi$ is a Dirac particle, this effective Lagrangian leads to the following $\chi N$ cross section \cite{Hill:2011be}

\begin{align}\label{Eq:DM-N_2}
 \sigma_{\chi N} = \frac{m_{\chi N}^2}{\pi \Lambda^6} \left| \sum_{q} \frac{C_q}{2 m_N} \langle N(p) |m_{q} \bar{q} q | N(p) \rangle  \right|^2,  
\end{align}

where $m_{\chi N}$ is the $\chi  N$ reduced mass. 
The matrix element in \eqref{Eq:DM-N_2} is a flavor separation of the sigma term, and is basically given by $\sigma_{\pi N}$ \cite{Crivellin:2013ipa}.
In dark matter literature, these couplings are given normally in terms of the effective couplings to quarks $f_{q}$,

\begin{align}\label{Eq:DM-N_3}
\frac{1}{2m_N}\langle N(p) |m_{q} \bar{q} q | N(p) \rangle = m_N f_{q}.
\end{align}

The cross sections for nuclear targets are written in terms of the effective couplings to protons and neutrons $f^{p,n}$, which are linear in $f_{q}$, as follows \cite{Hill:2011be},

\begin{align}\label{Eq:DM-N_4}
 \sigma_{\chi \mathcal{N}} = \frac{m_{\chi \mathcal{N}}^2}{\pi} \left| Z f^p + (A-Z) f^n  \right|^2,  
\end{align}

with $A$ the mass number of the nucleus $\mathcal{N}$ with charge $Z$, and $m_{\chi \mathcal{N}}$ the reduced mass of the $\chi \mathcal{N}$ system.

The interest for dark-matter detection is that the DM-nucleus cross sections depends quadratically on $f_q \propto \langle N(p) |m_{q} \bar{q} q | N(p) \rangle$, so the uncertainty in $\sigma_{\pi N}$ is magnified in $\sigma_{\chi \mathcal{N}}$ \cite{Bottino,Ellis:2008hf}. 
In the latter reference it is shown that, with the range of values available in the literature for the sigma term, the variation of the estimated spin-independent cross section could change in approximately a $50\%$. 
Given such a big impact, it is not surprising that the authors literally pleaded for an experimental campaign to determine better $\sigma_{\pi N}$, because ``{\it...it is potentially also a key ingredient in the effort to understand one of the
most important aspects of possible new physics beyond the standard model.}"

\subsection{Nucleosynthesis and the origin of life}

The $\pi N$ interaction is also an important part of the nuclear forces, since the latter inherit the properties from the former. 
The pion-nucleon sigma term, that is, the strength of the explicit chiral symmetry breaking of the strong force, is related to the intensity of the $\pi N$ interaction with scalar-isoscalar quantum numbers. 
This part of the interaction plays a prominent role in alpha-like particle clustering, as was shown using Nuclear Lattice EFT (NLEFT) \cite{Elhatisari:2016owd}.
Understanding this process becomes essential in order to explain the observed ${}^{12}\text{C}$ abundance, which is possible only though to the formation of a $0^+$ resonance near the ${}^4\text{He }\text{+ }{}^8\text{Be}$ threshold (the Hoyle state). 

Connected to this matter, in Ref.~\cite{Berengut:2013nh} the authors studied the quark mass dependence of nuclear binding energies and nuclear threshold parameters. 
The change in the nucleon mass, controlled by the value of $\sigma_{\pi N}$, is part of the input used in this study. 
From the primordial abundances of deuterium and helium they estimated a limit on the variation of the light quark masses and found a relative change of only $2(4)\%$. 

\subsection{Nuclear thermodynamics}

The matrix element of QCD related to the sigma term provides also valuable information about nuclear matter properties. 
When one studies the restoration of quiral symmetry in nuclear matter, one of the quantities used as order parameter in the phase transition is the quiral condensate in the nuclear medium, $\langle \Psi | \bar{q}q |\Psi \rangle$, where $|\Psi\rangle$ is the nuclear ground state \cite{Weise:2012yv}. 
Applying the Hellmann-Feynman theorem to the Hamiltonian density of QCD, $\mathcal{H}_{\text{QCD}}$, one gets

\begin{align}
 \langle \Psi | \bar{q}q |\Psi \rangle (\rho)=  \langle \Psi |\frac{\partial \mathcal{H}_{\text{QCD}} }{\partial m_q} | \Psi \rangle =  \langle 0 |\bar{q}q | 0 \rangle \left[ 1- \frac{1}{f_\pi^2}\frac{\partial \mathcal{E}(\rho,M_\pi)}{\partial M_\pi^2 }\right]
\end{align}

being $ \langle 0 |\bar{q}q | 0 \rangle$ the quark condensate in the vacuum, and $\mathcal{E}(\rho,m_\pi)$ the energy density of nuclear matter.
Deriving the latter, one finds \cite{Kaiser:2007nv}

\begin{align}\label{Eq:Condensate}
 \frac{\langle \Psi | \bar{q}q |\Psi \rangle}{\langle 0 | \bar{q}q | 0 \rangle}(\rho) = 1 - \frac{\rho}{f_\pi^2} \left\{ \frac{\sigma_{\pi N}}{M_\pi^2} \left( 1- \frac{3k_f^2}{10 m_N^2} + \frac{9 k_f^4}{56 m_N^4} \right) + D(k_f)  \right\} 
\end{align}

where $k_f$ is the Fermi momentum and $D(k_f)$ are the interaction contributions beyond the linear density approximation, calculable with the formalism developed in \cite{Oller:2001sn,Meissner:2001gz,Oller:2009zt,Lacour:2010ci,Lacour:2009ej,Oller:2019ssq}.

\begin{figure}[H]
 \begin{center}
 \includegraphics[width=.7\textwidth]{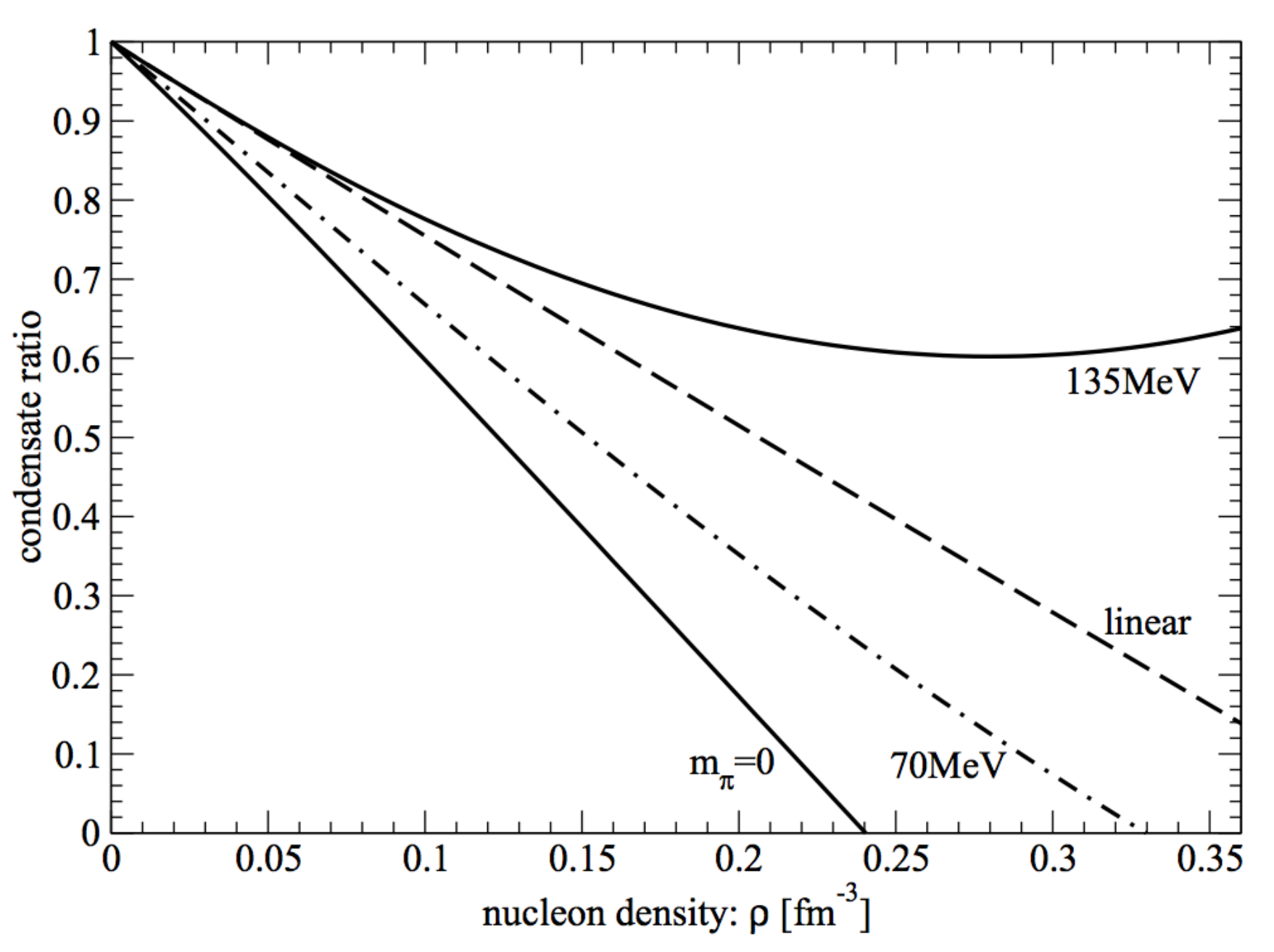} 
 \caption{The value of the ratio in Eq.~\eqref{Eq:Condensate} as a function of the density for three different values of the pion mass. 
 The dashed line corresponds to the value of $\sigma_{\pi N} = 45$~MeV assuming linear dependence in $\rho$. Reprinted figure with permission from N. Kaiser, P. de Homont, and W. Weise, Phys. Rev. C 77, 025204 (2008). Copyright (2008) by the
American Physical Society, \url{http://dx.doi.org/10.1103/PhysRevC.77.025204}.}
 \label{Fig:chiral_condensate}
 \end{center}
\end{figure} 

From the previous equation, we see that the sigma term controls the order parameter in the transition to a symmetric regime. 
In fact, it seems to indicate that, the bigger $\sigma_{\pi N}$ is, the lower is the density where this transition happens.
Fig.~\ref{Fig:chiral_condensate} shows explicitly this dependence in the linear density approximation for a sigma term of $45$~MeV. 
In that case, $\langle \Psi | \bar{q}q |\Psi \rangle = 0$ for $\rho \approx 3 \rho_0$, where $\rho_0$ is the nuclear saturation density. 
If one considers $\sigma_{\pi N} = 60$~MeV, $\langle \Psi | \bar{q}q |\Psi \rangle$ would vanish at $\rho \approx 2 \rho_0$.

However, it is important to consider too, that in order to have a spontaneous breaking of the chiral symmetry, it is also necessary to have a non-vanishing temporal component of the pion axial coupling $f_t$ \cite{Meissner:2001gz,Alarcon:2011zs}.
Therefore, studying the recovery of chiral symmetry in nuclear matter requires additional analysis.

\subsection{Strangeness content of the nucleon}

In the same way as $\langle N |  \hat{m}(\bar{u}u + \bar{d}d )| N \rangle$ is relevant in different aspects of particle and nuclear physics, is equally relevant for these fields to study $\langle N | m_s \bar{s}s| N \rangle$. 
The later, is related to the contribution of the strange quark to the nucleon mass. 
What is interesting is that this matrix element can be related to $\sigma_{\pi N}$ though group-theoretical arguments. 

Let's begin by considering the mass term of the QCD Hamiltonian,

\begin{align}
 \mathcal{H}_m  = m_u \bar{u}u + m_d \bar{d}d + m_s \bar{s}s
\end{align}

that can be written as,

\begin{align}
 \mathcal{H}_m  = m_u \bar{u}u + m_d \bar{d}d + m_s \bar{s}s = \frac{1}{3}(m_s + 2 \hat{m})( \bar{u}u + \bar{d}d + \bar{s}s ) + \frac{1}{3}(\hat{m} - m_s)( \bar{u}u + \bar{d}d -2 \bar{s}s ).
\end{align}

The point is that the second term contributes differently to the baryon octet masses, and therefore the matrix element $\langle N |( \bar{u}u + \bar{d}d -2 \bar{s}s ) |N \rangle$ can be related to the mass splitting of these baryons at leading order in the $SU(3)$-breaking parameter $m_s- \hat{m}$ \cite{Alarcon:2012nr,Fernando:2018jrz},

\begin{align}\label{Eq:sigma0}
 \sigma_0 \equiv \frac{\hat{m}}{2 m_N} \langle N |( \bar{u}u + \bar{d}d -2 \bar{s}s )|N \rangle = \frac{\hat{m}}{m_s - \hat{m}} (m_\Xi + m_\Sigma - 2 m_N)
\end{align}

where $\sigma_0$ has been defined in analogy with $\sigma_{\pi N}$.

The relation of $\sigma_0$ with the octet baryon mass splitting provides a way of estimating the matrix element in $\sigma_s$, assuming a known value of $\sigma_{\pi N}$.
From the definitions of $\sigma_{\pi N}$, $\sigma_s$ and $\sigma_0$ is easy to obtain

\begin{align}
 \sigma_s = \frac{m_s}{2\hat{m}} (\sigma_{\pi N} - \sigma_0)
\end{align}

once $\sigma_{\pi N}$ and $\sigma_0$ are known.
Notice that if one assumes a small $\langle N | \bar{s}s | N \rangle$ (Zweig rule), the value of the sigma term should be well approximated by $\sigma_0$  \cite{Cheng:1975wm}. 
If the relation of the latter with the baryon masses, Eq.~\eqref{Eq:sigma0}, is accurate enough, then the sigma term should be around $30$~MeV.
Values of $\sigma_{\pi N} \approx 60$~MeV seem to imply a huge value of  $\sigma_s$, given the quark mass ratio value $m_s/2\hat{m}\approx 13$.
Historically, the relation of $\sigma_{\pi N}$ with $\sigma_s$ has played a crucial role in the long-standing controversy related to the value of $\sigma_{\pi N}$, as I will explain in more detail in Sec.~\ref{Sec:sigma-term_puzzle}.

\section{Methods to determine the sigma term}

In this section I will comment very briefly on some of the different approaches and strategies used to determine the sigma term. 
This will allow to understand better the controversy on its value that will be presented in the next section.

\subsection{Models of $SU(3)$ breaking}
\label{Sec:ModelsSU3Breaking}

After the discovery of the approximate invariance of the strong interactions under $SU(3)$ \cite{Neeman:1961jhl,GellMann:1962xb} and the suggestion of approximate symmetry under the chiral group $SU(3)\times SU(3)$ \cite{GellMann:1964tf}, many models for chiral symmetry breaking have been investigated in different contexts of the strong interactions. 

In the case of the pion-nucleon sigma term, it was customary in the 70s to propose a specific representation of the $SU(3)\times SU(3)$ symmetry group, which usually was the $(3,\bar{3})\oplus(\bar{3},3)$ \cite{GellMann:1968rz,VonHippel:1970uz},

\begin{align}
 \mathcal{H}_{SB} = - c_0 u_0 - c_8 u_8,
\end{align}

where $u_0$ and $u_8$ are the singlet and octet members of the $SU(3)$ nonet of scalar densities, and $c_0$ and $c_8$ are numerical coefficients. 
Since $u_8$ is an octet, the coefficient $c_8$ could be easily determined through octet mass splittings. 
However, the singlet part became problematic given that, in principle, there was no clear evidence of how large could be the singlet contribution to the nucleon mass. 
 Some works tried to face this problem by estimating this contribution though the baryon octet mass average \cite{VonHippel:1970uz}, but it was difficult to figure out how reliable this estimation was.

Few years later, Cheng made an important observation. He pointed out in \cite{Cheng:1975wm} that, if one assumes $\langle N | \bar{s}s | N \rangle  \ll \langle N | \bar{u}u + \bar{d}d | N \rangle $ (Zweig rule), the sigma term could be determined by the octet mass splitting, 

\begin{align}
\sigma_{\pi N} = \frac{1}{2m_N}\langle  N |\hat{m} (\bar{u}u + \bar{d}d)| N \rangle \approx \frac{1}{2m_N}\langle  N |\hat{m} (\bar{u}u + \bar{d}d - 2 \bar{s}s)| N \rangle. 
\end{align}

Since the last term is purely octet, it can be related to the mass difference, so that

\begin{align}\label{Eq:II.A-3}
\sigma_{\pi N} \approx \frac{\hat{m}}{m_s - \hat{m}} (m_\Xi + m_\Sigma - 2 m_N) . % \approx 27~\text{MeV}. 
\end{align}

This approach allows to relate the sigma term to the hadron masses {\it assuming} a negligible contribution of $\bar{s}s$ compared to $\bar{u}u + \bar{d}d$ {\it and} the validity of 

\begin{align}\label{Eq:II.A-4}
\frac{1}{2m_N}\langle  N |\hat{m} (\bar{u}u + \bar{d}d - 2 \bar{s}s)| N \rangle =  \frac{\hat{m}}{m_s - \hat{m}} (m_\Xi + m_\Sigma - 2 m_N). 
\end{align}

The latter is obtained from a leading-order $SU(3)$-breaking calculation, that is expectable to work given the success of the Gell-Mann-Okubo (GMO) mass formula $\Delta_{GMO} = 3m_\Lambda+m_\Sigma - 2(m_N+m_\Xi)$. 
I will show later that relatively large corrections to Eq.~\eqref{Eq:II.A-4} do not spoil the good accuracy of $\Delta_{GMO}$.

\subsection{Dispersion Relations} 

Dispersion theory is among the first methods used to estimate the value of the sigma term. 
The starting point is the standard decomposition of the $\pi N$ scattering amplitude in terms of the four Lorentz invariant amplitudes $A^\pm$ and $B^\pm$, or $D^\pm$ and $B^\pm$  [see Eq.~\eqref{Eq:T-decomposition}].
One can write a dispersion relation to reconstruct $\bar{D}^+$ from $\pi N$ scattering data and evaluate this amplitude at the Cheng-Dashen point to extract $\sigma_{\pi N}$. In terms of the pion lab energy $\omega \equiv \frac{s-m_N^2 - M_\pi^2 }{2 m_N}$ this relation reads \cite{Gasser:1988jt},

\begin{align}
  \bar{D}^+ (\omega = - \frac{M_\pi^2}{2 m_N}) =  \bar{D}^+ (\omega = M_\pi^2) +  \frac{2( \frac{M_\pi^4}{4 m_N^2})}{\pi} \int_{M_\pi}^\infty \frac{d\omega' }{\omega'^2- \frac{M_\pi^4}{4 m_N^2}}\frac{\text{Im}D^+(\omega')}{\omega'^2 - M_\pi^2}, 
\end{align}

where $\text{Im}D^+(\omega)$ is directly related to the scattering data.
%Another application of dispersion theory was proposed by Fubini and Furlan [S. Fubini and G. Furlan, Ann. Phys. (N. Y.) 48, 322 (1968)] berfore the article by Cheng and Dashen [XXX]. They propose a method to extrapolate the "soft-theorem" results derived from current algebra (valid for $M_\pi = 0$) to the physical point using dispersion relations in the mass variable. 
However, this calculation requires the knowledge of $\Delta_\sigma$ (see Eq.~\eqref{Eq:Pagels&Pardee} and text below). 
For that, one needs to consider the scalar form factor of the nucleon 

\begin{align}
 \sigma(t)  = \frac{1}{2m_N}\langle  N(p') |\hat{m} (\bar{u}u + \bar{d}d)| N(p) \rangle, \hspace{1cm} \hat{m} = \frac{m_u + m_d}{2},  \hspace{1cm}  t=(p'-p)^2.
\end{align}

One can write a dispersive representation for $\sigma(t)$ as was done in \cite{Gasser:1990ap}

\begin{align}
 \sigma(t) = \sigma_{\pi N} + \frac{t}{\pi} \int_{4M_\pi^2}^\infty dt' \frac{\text{Im} \sigma(t')}{t' (t' - t - i\epsilon)}   
\end{align}

where $\sigma_{\pi N} = \sigma(0)$. 
Through this representation is clear the importance of the $\pi\pi$ interaction in this dispersive evaluation, since the imaginary part of $\sigma(t)$ is related to the scalar form factor of the pion $\Gamma_\pi(t)$,

\begin{align}
 \text{Im}\sigma(t) = \frac{3}{2} \frac{\Gamma_\pi^*(t) f_+^0(t)}{4m_N^2 - t }\left( 1 - \frac{4 M_\pi^2}{t} \right)^{1/2}
\end{align}

where $f_0^+(t)$ is the $I = J = 0$ $\pi \pi \to \bar{N}N$ partial wave. The pion scalar form factor $\Gamma_\pi(t)$ can be calculated in terms of the $\pi \pi$ phase shifts using an Omn\`es representation \cite{Oller:2007xd} or by solving the associated Muskhelishvily-Omn\`es problem in coupled channels \cite{Donoghue:1990xh}. However, this dependence on the $\pi \pi$ phase shifts was, in principle, not desirable, given the difficulty to have a reliable determination at that time \cite{Gasser:1990ce}. In this sense, the previous work achieved a major progress in the dispersive evaluations of the sigma term, since they designed a method that reduces this dependence significantly.

The starting point is to consider an expansion of $\bar{D}^+ (\nu, t)$ around the point $\nu = 0, t = 0$, where $\bar{D}^+ (\nu, t)$ is real,

\begin{align}
 \bar{D}^+ (\nu, t) = \bar{d}_{00}^+  + \bar{d}_{10}^+ \nu^2 + \bar{d}_{01}^+ t + \dots
\end{align}

Evaluating this expansion at the Cheng-Dashen point, one gets

\begin{align}\label{Eq:30}
\Sigma = f_\pi^2 \bar{D}^+ (0, 2M_\pi^2) = f_\pi^2(\bar{d}_{00}^+  + 2 \bar{d}_{01}^+ M_\pi^2) + \Delta_D
\end{align}

where $\Delta_D = \bar{d}_{02} t^2 + \dots$ is called the curvature term. 
Since, according to the (modified) Cheng-Dashen theorem, 

\begin{align}\label{Eq:31}
\Sigma = f_\pi^2 \bar{D}^+ (0, 2M_\pi^2) = \sigma(t = 2M_\pi^2) + \Delta_R = \sigma_{\pi N} + \Delta_\sigma + \Delta_R,
\end{align}

the comparison of Eqs.~\eqref{Eq:30} and \eqref{Eq:31} gives the following expression for the sigma term 

\begin{align}\label{Eq:GLS1}
  \sigma_{\pi N} = f_\pi^2(d_{00}^+  + 2 d_{01}^+ M_\pi^2) + \Delta_D - \Delta_\sigma - \Delta_R.
\end{align}

The crucial observation done in \cite{Gasser:1990ce} was that, in the evaluation of the difference $\Delta_D - \Delta_\sigma$, the effect of the $\pi\pi$ intermediate states almost cancel, giving a small contribution, $\Delta_D - \Delta_\sigma = -3.3(2)$~MeV. 
Then, the evaluation of  $\sigma_{\pi N}$ depends mostly on the values of subthreshold coefficients $d_{00}^+$ and $d_{01}^+$, since $\Delta_R$ is known to be small \cite{Bernard:1996nu}.  
The cancellation of the $\pi\pi$ intermediate state in the evaluation of the sigma term found in this work will be also relevant for evaluations of  $\sigma_{\pi N}$ using effective field theories, as I will comment later.

Apart from the fixed-$t$ approach, there are other calculations that do not rely on fixed-$t$ dispersion relations, as the hyperbolic dispersion relations, that made important contributions to the determination of the sigma term as well. 
A remarkable example is the determination of $\sigma_{\pi N}$ using Roy-Steiner equations \cite{Hoferichter:2015dsa}. 
They are a set of coupled partial-wave dispersion relations that implement the constrains from unitarity, analyticity and crossing symmetry in the $s$- and $t$-channels of $\pi N \to \pi N$.
The solution of the $t$-channel requires input from $\pi \pi \to \pi \pi$ in the form of phase shifts, which is usually taken from other works.
The results appear in the form of phase shifts for $\pi N \to \pi N$, which are obtained by solving numerically an integral equation.
These equations were proposed for $\pi N$ scattering in Ref.~\cite{Becher:1999he} in the context of a chiral effective field theory (chiral EFT) calculation using Infrared Regularization, but no explicit solution was given. 
In Ref. \cite{Hoferichter:2015dsa}, the authors followed the strategy of Ref. \cite{Gasser:1990ce}, and determined $\sigma_{\pi N}$ via the subthreshold coefficients $\bar{d}^+_{00}$ and $\bar{d}^+_{01}$, including a leading order estimation of the isospin breaking effects with chiral EFT. 
In their calculation they also studied the cancellation of $\Delta_D$ and $\Delta_\sigma$ and they found $\Delta_D - \Delta_\sigma = 1.8(2)$~MeV, what confirmed the findings of \cite{Gasser:1990ce}.

\subsection{Sum Rules}

Sum rules can be seen as a version of dispersive determination of the sigma term.  
Continuing with the work of Ref.~\cite{Gasser:1988jt}, Eq.~\eqref{Eq:GLS1} can be expressed in terms of the threshold parameters in $\pi N$ scattering to end up with a direct relation between these quantities and $\sigma_{\pi N}$.
These kind of relations started to appear in the literature in the early 70s \cite{Altarelli:1971kh}, and continued later with improved versions \cite{Olsson:1979ee,Olsson:1999jt}. 

What is interesting in the sum rules determinations, is that they allow us to see the weight of the different contributions to the final value of the sigma term. Taking as example the sum rule derived by Olsson in \cite{Olsson:1999jt}

\begin{align}
 f_\pi^{-2}\Sigma &= 14.5 a_{0+}^+ - 5.06 \left( a_{0+}^{(1/2)}\right)^2 - 10.13 \left( a_{0+}^{(3/2)}\right)^2 - 5.55 C^+ - 0.06 a_{1-}^{(+)} \nonumber \\
                          &+ 5.70 a_{1+}^+ - (0.08\pm 0.03)\hspace{4cm}  [\text{in $M_\pi^{-1}$}]
\end{align}

it is clear that the threshold parameters $a_{0+}^+$ and $a_{0+}^{(3/2)}$ are the most relevant ones in the actual value of the sigma term. 
In fact, part of the controversy on the size of the sigma term can be attributed to the difficulty of determining accurately $a_{0+}^+$. 
In the recent years, with the improvements in spectroscopy experiments of pionic deuterium and pionic hydrogen atoms \cite{Gotta:2008zza}, we had the chance for a more precise determination of $a_{0+}^+$ and $a_{0+}^{(3/2)}$ \cite{Baru:2011bw}. 
As I will discuss later, this extraction is the key to solve the discrepancy between the different phenomenological determinations of the sigma term, given the strong correlation of these quantities (specially $a_{0+}^+$) to the value of $\sigma_{\pi N}$.

\subsection{Chiral Effective Field Theory}

Chiral EFT is an approach suited to study the strong interactions at low energies. 
It opens the possibility to determine systematically the low energy structure of the QCD green functions \cite{Gasser&Leutwyler}, and provide model-independent results for hadronic processes and hadron properties at low energies. 
The fact that chiral EFT is constructed to systematically provide corrections to the low-energy theorems of QCD, makes it a natural tool to calculate the pion-nucleon sigma term. 
In Chiral EFT one can provide a systematic low-energy expansion of a QCD operator in terms of the low-energy degrees of freedom, that can be calculated analytically with controlled accuracy. 
Therefore, in an EFT calculation is not mandatory to go through the Cheng-Dashen theorem, which can be seen as a way to connect experimental information ($\pi N$ scattering amplitude) at an unphysical point to a QCD matrix element. 
Although in chiral EFT there is no problem in using the results of \cite{Cheng:1970mx}, since preserves the analytic structure of the amplitude\footnote{Only in a relativistic formulation of chiral EFT with baryons the analytic structure is completely preserved. However, since the problematic parts in non-relativistic approaches are the Born terms, and they are subtracted in the Cheng-Dashen theorem, one can still apply it in these versions of the EFT.}. The leading low-energy operator that determines the value of the sigma term is

\begin{align}\label{Eq:LEC_c1}
 \mathcal{L}_{\pi N}^{(2)} 	\supset c_1 \langle \chi_+ \rangle \bar{\psi}\psi  
\end{align}

where $\psi$ refers to the nucleon isodoublet, and  $\langle \chi_+ \rangle$ is a trace in the isospin space of a matrix involving even powers of the pion field and quark masses (see Ref.~\cite{Alarcon:2012kn} for more details).

In chiral EFT one can determine $\langle N|\hat{m}( \bar{u}u + \bar{d}d )|N\rangle$ by calculating the scalar form factor at $t=0$ using the low-energy chiral representation of the scalar operator in QCD, or by applying the Hellmann-Feynman theorem \cite{Hellmann-Feynman} to the chiral representation of the nucleon mass since. According to this theorem, the sigma term is related to the nucleon mass in the following way,\footnote{The last equality is valid only in a two-flavor approach.}

\begin{align}
 \sigma_{\pi N} = \hat{m} \frac{\partial m_N}{\partial \hat{m}} =  M_\pi^2 \frac{\partial m_N}{\partial M_\pi^2}. 
\end{align}

Nevertheless, the crucial point in chiral calculations is the determination of the low-energy constants (LECs) that parametrize the low-energy expansion of the QCD operator.
Whether they are determined from fits to data or from other observables, it does not guarantee a reliable parametrization. 
The reason is that, in the end, one is matching an observable to a perturbative expansion of the QCD Green function, and needs to study the convergence of the calculation. 
Therefore the determination of the low-energy constant has to be tested to prove the reliability of the determination of $\sigma_{\pi N}$.

Fortunately, chiral symmetry relates the value of these low-energy constants to different low-energy processes, so one can use a given observable to determine the necessary LECs and make predictions for other different observables to test the extracted values. 
Following this strategy one also finds that the value of $\sigma_{\pi N}$ is involved in other processes, and this gives a broader view of the importance of the QCD matrix element  $\langle N|\hat{m}( \bar{u}u + \bar{d}d )|N\rangle$.

\begin{figure}[H]
 \begin{center}
 \includegraphics[width=.9\textwidth]{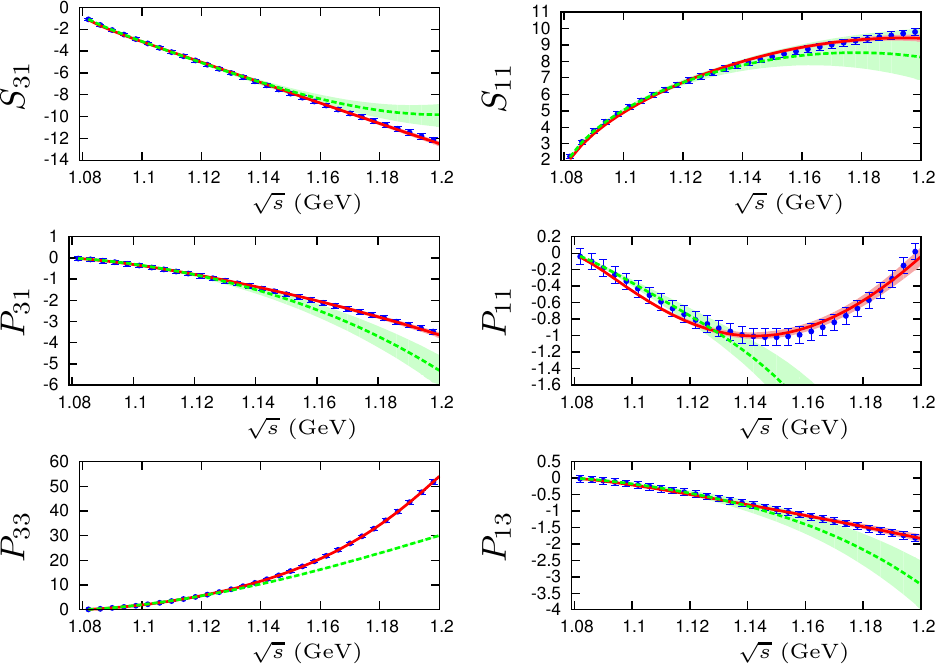} 
 \caption{Example of chiral fits to $\pi N$ phase shifts. The red band corresponds to a relativistic calculation with explicit $\Delta$, while the green one is without the $\Delta$. Reprinted from Annals of Physics, Volume 336, J. M. Alarc\'on, J. Mart\'in Camalich and J. A. Oller, ``Improved description of the $\pi N$-scattering phenomenology at low energies in covariant baryon chiral perturbation theory", 413-461, Copyright (2013), with permission from Elsevier, \url{https://doi.org/10.1016/j.aop.2013.06.001}. }
 \label{Fig:ChiralEFT_fits}
 \end{center}
\end{figure} 

\subsection{Lattice QCD}

Lattice QCD is a numerical method to compute QCD Green functions. 
Since this approach allows to calculate QCD matrix elements non-perturbatively, it becomes one of the most interesting way to compute the sigma term. As in the EFT approach, one can follow, at least, two different strategies. 
The first one is to calculate the sigma term through the Hellmann-Feynman theorem. 
For that, one needs the quark mass dependence of the nucleon mass. 
In lattice QCD this dependence is computed from the two-point functions, with two nucleon interpolating fields in different space-time positions

\begin{align}\label{Eq:2-point_correlator}
 C_N (t_s)\propto \sum_{\vec{x}_s}\langle J_N(\vec{x}_s,t_s)\bar{J}_N(\vec{x}_0,t_0) \rangle,
\end{align}

where $C_N$ is the zero-momentum two-point correlator of the nucleon, $\vec{x}_s$ and $\vec{x}_0$ are the sink and source coordinates and $t_s$ and $t_0$ their corresponding times.\footnote{In this section, $t$ refers always to time.} $J_N$ is the nucleon interpolating field, and for the proton is given by

\begin{align}
 J_p  = \epsilon_{abc}\left( u_a^T C \gamma_5 d_b \right)u_c, 
\end{align}

with $u$ and $d$ the quark fields and $C$ the charge conjugation matrix.
Since the behavior of Eq.~\eqref{Eq:2-point_correlator} at large times is dictated by the lowest mass state with the same quantum numbers as the current, the nucleon mass is extracted by studying the behavior of the correlator with the time

\begin{align}
 M_{\text{eff}} = -\log\left(\frac{C_N(t)}{C_N(t-1)} \right),
\end{align}

where the effective mass ($ M_{\text{eff}}$) is identified with the lowest mass state for large enough $t$.
By repeating this procedure for different quark masses one can parametrize quark mass dependence of the nucleon mass needed to apply the Hellmann-Feynman Theorem.
The functional form of this parametrization, needed to take the derivative with respect to the quark mass, is external to the lattice calculation.
The typical quark mass dependences proposed in the literature are polynomials or chiral expansions, being the former the most popular ones. 
However, it is important to stress that polynomial (or Pad\'e) parametrizations do not provide the expected logarithm dependence on the quark masses predicted by the low-energy theorems of QCD.

The second possibility is to calculate the sigma term as the scalar form factor of the nucleon at $q^2 = 0$. 
This type of calculation requires the computation of three point functions, and therefore becomes more computationally expensive than the previous one. 
Nevertheless, it has the important advantage that it does not involve any parametrization as in the Hellmann-Feynman type. 
The procedure is similar to the two-point function type of calculation.
The difference is that here one studies the evolution with time time of the ratio of the three-point function (insertion of the scalar operator $\mathcal{O}$) to the two-point function \cite{Abdel-Rehim:2016won},

\begin{align}\label{Eq:Lattice_FF}
R(t_s;t_{\text{ins}}) &= \frac{C^{\mathcal{O}}_{\text{3pt}}(t_s - t_0, t_{\text{ins}}- t_0) }{C_{\text{2pt}}(t_s - t_0)} \nonumber \\
                               &\propto \frac{\sum_{\vec{x}_s,\vec{x}_{\text{ins}}} e^{-i (\vec{x}_s - \vec{x}_0)\cdot \vec{p}}  \langle J_N(\vec{x}_s,t_s)\mathcal{O}(\vec{x}_\text{ins},t_\text{ins})\bar{J}_N(\vec{x}_0,t_0) \rangle}{\sum_{\vec{x}_s}\langle J_N(\vec{x}_s,t_s)\bar{J}_N(\vec{x}_0,t_0) \rangle}
\end{align}

with $C^{\mathcal{O}}_{\text{3pt}}$ and $C_{\text{2pt}}$ the three- and two-point functions. 
They depend on the insertion-source time separation $t_{\text{ins}}-t_0$ and the sink-insertion separation, $t_s - t_{\text{ins}}$. 
These calculations exploit the fact that when $t_{\text{ins}}-t_0$ and $t_0 - t_{\text{ins}}$ are large enough, the contributions to the matrix element of $\mathcal{O} = m_q \bar{q}q$ from the exited states are negligible. 
Inserting a complete set of intermediate states in Eq.~\eqref{Eq:Lattice_FF}, the ratio takes the form 

\begin{align}\label{Eq:Lattice_FF_2}
R(t_s;t_{\text{ins}}) \propto \frac{\sum_{i,j} \langle J |\phi_j\rangle \langle\phi_i | \bar{J}\rangle \langle \phi_j|\mathcal{O}|\phi_i \rangle e^{-E_i (t_{\text{ins}}-t_0 )}e^{-E_{j} (t_s-t_{\text{ins}} )}}{\sum_{i}|\langle J |\phi_i\rangle|^2 e^{-E_i (t_s-t_0 )}}
\end{align}

where $| \phi_i \rangle$ is the $i^{\text{th}}$ eigenstate of the QCD Hamiltonian with the nucleon quantum numbers, and $E_j$ is the energy of the state in the rest frame. Note that the two-point function in the denominator is introduced to cancel the unknown overlap of the source $J$ with the state $\phi_i$. 
If one defines $\delta E_i = E_i - E_0$, being $E_0$ the ground state energy, it is simple to see that when $\delta E_i(t_s - t_{\text{ins}}) \gg 1$ and $\delta E_i(t_{\text{ins}} - t_0) \gg 1$ 

\begin{align}\label{Eq:Lattice_FF_3}
R(t_s;t_{\text{ins}}) \to \langle \phi_0 | \mathcal{O}|\phi_0 \rangle = m_q \langle N |\bar{q}q|N \rangle
\end{align}

\begin{figure}[H]
 \begin{center}
 \includegraphics[width=.49\textwidth]{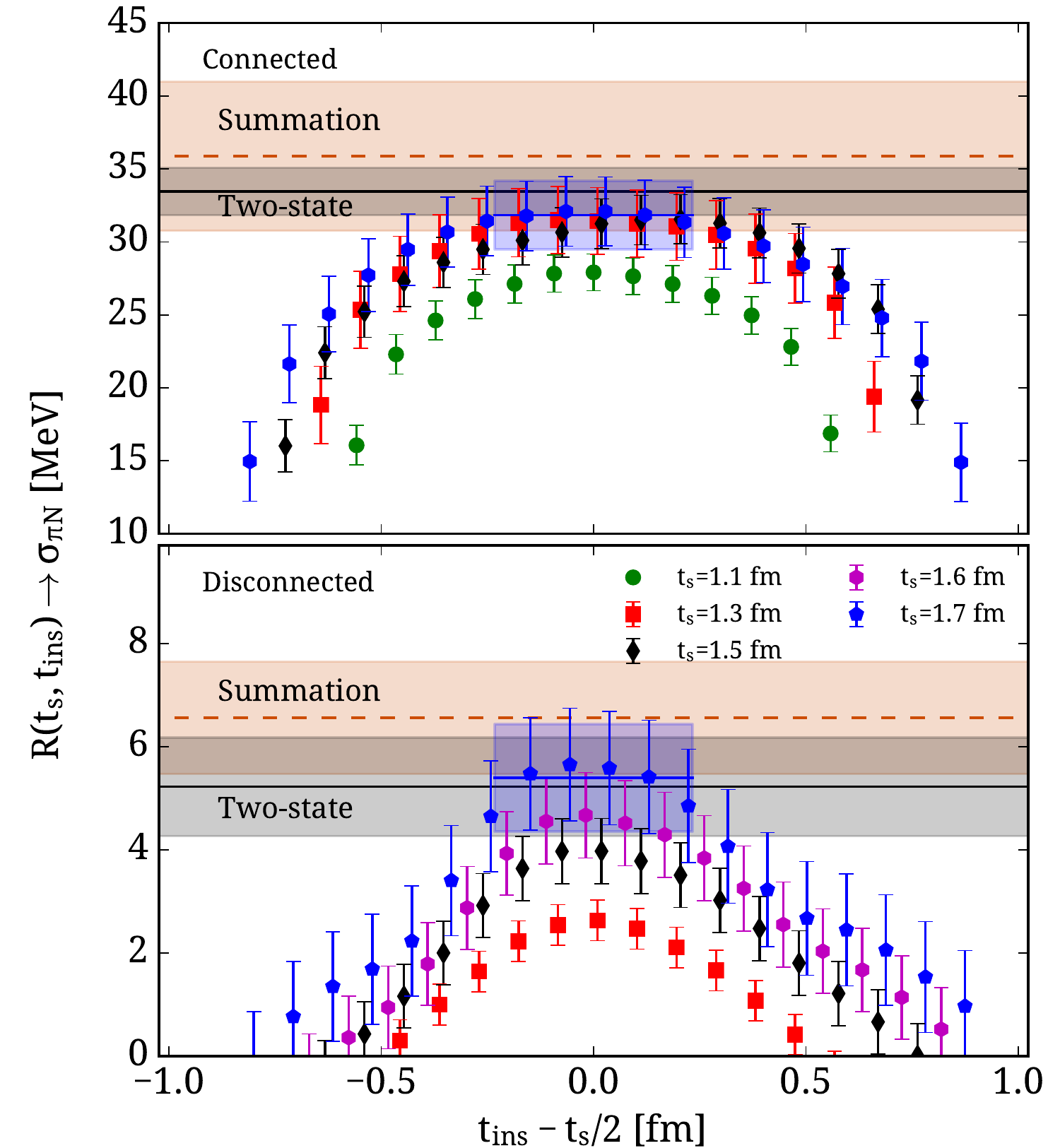} 
 \includegraphics[width=.49\textwidth]{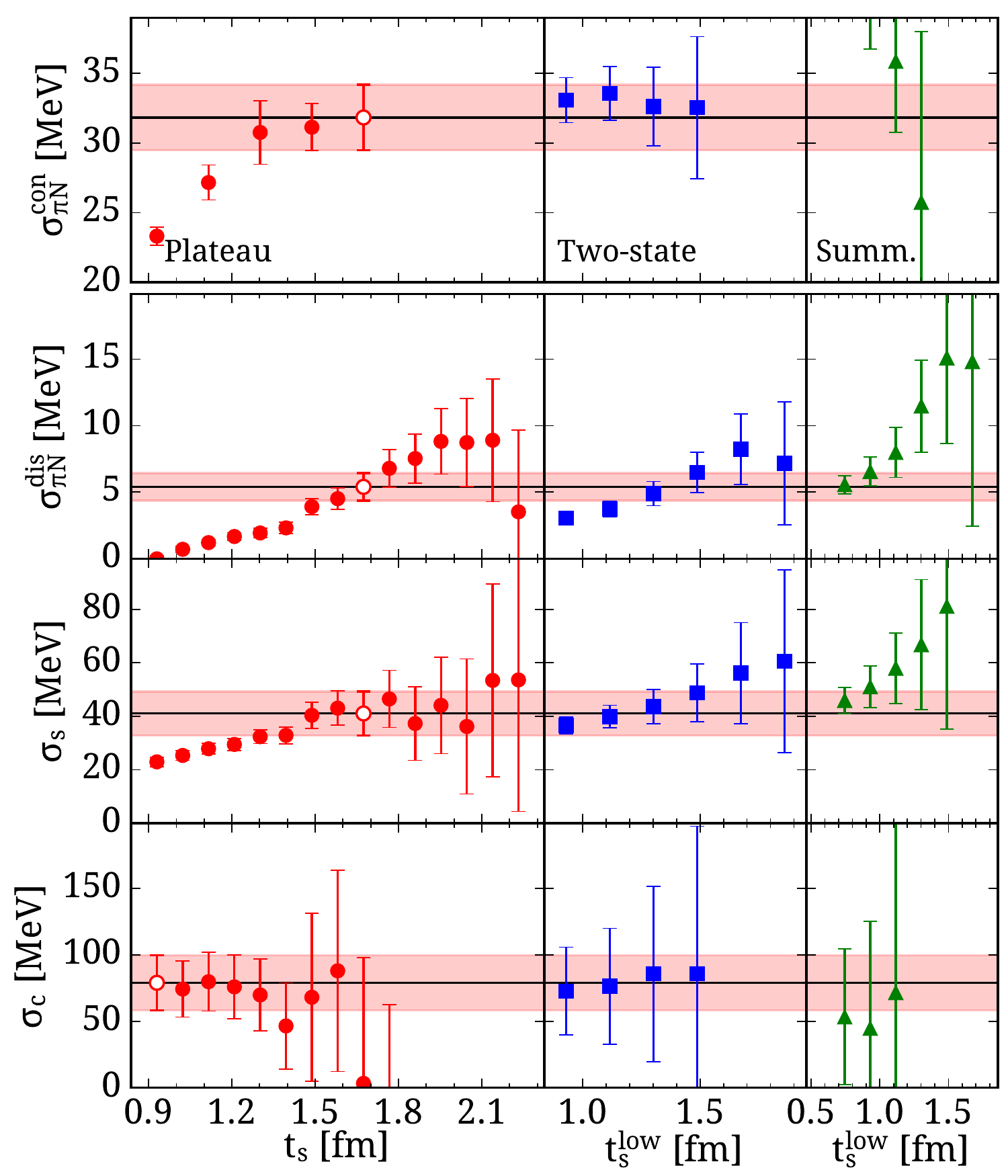} 
 \caption{Example of a lattice extraction of $\sigma_{\pi N}$ computing the three-point function. $R(t_{\text{s}},t_{\text{ins}})$ is the ratio of the three-point function to the two-point one. The ratio approaches to $\sigma_{\pi N}$ when $t_{\text{s}} - t_{\text{ins}}$ (sink-insertion separation) and $t_{\text{ins}}-t_{0}$ (insertion-source separation) are large enough (see Ref.~\cite{Abdel-Rehim:2016won} for more details). Reprinted figure with permission from A. Abdel-Rehim, C. Alexandrou, M. Constantinou, K. Hadjiyiannakou, K. Jansen, Ch. Kallidonis, G. Koutsou, and A. Vaquero Avil\'es-Casco, Phys. Rev. Lett. 116, 252001. Copyright (2016) by the
American Physical Society, \url{http://dx.doi.org/10.1103/PhysRevLett.116.252001}.}
 \label{Fig:Lattice}
 \end{center}
\end{figure}

Lattice calculations of the sigma term have also some difficulties to face. 
One of them, affecting the Hellmann-Feynman and three-point function methods, is the possible contamination of the excited states in the computation of $m_N(m_q)$ and $\sigma_{\pi N}$.
With time, several strategies that complement the usual {\it plateau} method have been developed in order to reduce the possible systematics of the extraction.
Fig.~\ref{Fig:Lattice} shows an example of extraction of $\sigma_{\pi N}$ with the three-point function method, and using additional strategies (two-states and summation) to avoid excited states contamination. 
Another important difficulty affecting only the calculation of the three-point functions is the contribution of the sea quarks (disconnected diagrams) to the matrix element.
The limitation comes from the computational effort needed for the calculation of these contributions. 
Since they involve the evaluation of a closed quark loop, it requires to know the quark propagator from all to all spatial coordinates, which is much more computationally demanding than the connected contributions. 
In the last years this contribution to the sigma term has been calculated at the physical point and is estimated to be at the $10\%$ level. 

Although the first calculations in lattice QCD were not able to compute the $\sigma_{\pi N}$ at the physical point, nowadays there are various results in the literature for physical pion mass obtained by different groups and with different methods \cite{Durr:2015dna,Yang:2015uis,Abdel-Rehim:2016won,Bali:2016lvx,Alexandrou:2017qyt,Borsanyi:2020bpd}. 
As I will comment below, these results at the physical point played a very important role in the current status of the sigma term.

\subsection{Pionic atoms}

As mentioned in Sec.~\ref{Sec2} the value of the sigma term has consequences for nuclear matter calculations. 
This fact was exploited in \cite{Friedman:2019zhc} to give an estimation of the sigma term based on pionic atoms results. 
The strategy of this approach is to relate the observed modification of the isovector scattering length $a_{0+}^-$ in pionic atoms to $\sigma_{\pi N}$. 
They construct the relation from the leading order chiral result for $a_{0+}^-$ in vacuum, given by the Weinberg-Tomozawa term

\begin{align}
a_{0+}^- = - \frac{M_\pi m_N}{8\pi(m_N + M_\pi)f_\pi^2 } .
\end{align}

One can study the in-medium modification of $a_{0+}^-$ by introducing the density dependence of $f_\pi$ [see Eq.~\eqref{Eq:Condensate}]

\begin{align}
\frac{f_\pi^2(\rho)}{f_\pi^2} = \frac{\langle \Psi | \bar{q}q |\Psi \rangle}{\langle 0 | \bar{q}q | 0 \rangle}(\rho) \approx 1 - \rho  \frac{\sigma_{\pi N}}{f_\pi^2 M_\pi^2} 
\end{align}

what modifies the vacuum result in the following way,

\begin{align}
 \hat{a}_{0+}^- = a_{0+}^- \left( 1- \rho\frac{\sigma_{\pi N}}{f_\pi^2 M_\pi^2 } \right)^{-1}
\end{align}

where $\hat{a}_{0+}^-$ refers to the in-medium value of $a_{0+}^-$.
In-medium modifications of $M_\pi$ can be considered as well, but it does not introduce substantial changes in the extraction of $\sigma_{\pi N}$ \cite{Friedman:2019zhc}.

This method shows one interesting difference with respect to the other phenomenological approaches: 
It is mostly based on the value of the isovector scattering length, which extraction is less model-dependent than the isoscalar one, used commonly in dispersion theory or EFT calculations. 
Therefore, it offers a determination of the sigma term completely independent form the previous methods.

\newpage
\section{The sigma term puzzle}
\label{Sec:sigma-term_puzzle}

{\it The early calculations and the beginning of the puzzle}
\\

The first determinations of the pion-nucleon sigma term appeared along with the investigation of different patterns of chiral $SU(3)\times SU(3)$ breaking, time before QCD was established as the fundamental theory of the strong interactions.
As explained in Sec.~\ref{Sec:ModelsSU3Breaking}, the most popular one was the $(3,\bar{3})\oplus(\bar{3},3)$ model, that leads (under certain assumptions) to $\sigma_{\pi N} \sim 30$~MeV \cite{VonHippel:1970uz}. 
This changed drastically after the work of Cheng and Dashen \cite{Cheng:1970mx}, that gave the possibility of connecting experimental information at one unphysical point (Cheng-Dashen point) to the value of the sigma term. 
In their original work, Cheng and Dashen used fixed-$t$ dispersion relations to relate the existing results of $\pi N$ phase-shifts to the value of $\sigma_{\pi N}$.
They got $\sigma_{\pi N} = 110$~MeV, a value much larger than the $SU(3)\times SU(3)$ breaking models.
This ``serious disagreement" (in words of Cheng and Dashen) was the start of a controversy on the value of $\sigma_{\pi N}$ that has been recurring until today.
At the beginning, the discrepancy between the different extractions was not so dramatic. 
Both, the models for $SU(3)\times SU(3)$ breaking and the dispersive evaluations had important limitations. 
The former, relied on assumptions on the singlet contribution to the nucleon mass, what introduced uncontrolled systematic error to the calculation. 
In the case of the dispersive results, the experimental input or the method used to reach the Cheng-Dashen point were questioned. 
In fact, not all the dispersive extractions agreed on a large value of $\sigma_{\pi N}$.
For example, H\"ohler and collaborators obtained a much lower value, $\sigma_{\pi N} = 42 (10)$~MeV \cite{Hoehler:1971km,Hoehler:1972gb},  
using fixed-$t$ dispersion relations, but with a different subtraction method. 
Later dispersive results usually got a relatively large value, but not as large $\sim 100$~MeV.
Examples of these calculations are Refs.~\cite{Nielsen:1974qz,Hite:1974wz,Langbein:1974rz,Chao:1975sk}, that obtained values from $57$--$68$~MeV for the sigma term.
\\

{\it The strangeness puzzle}
\\

In the mid 70s it was assumed that dispersion theory calculations obtained values around $60$~MeV for $\sigma_{\pi N}$, and with time they became more accepted than the results from models for $SU(3)\times SU(3)$ breaking. 
This situation changed drastically after publication of \cite{Cheng:1975wm}, where the community noticed a problem with the dispersive (large) values of $\sigma_{\pi N}$. 
The crucial observation of this work is that, assuming $\langle N | \bar{s}s | N \rangle  \ll \langle N | \bar{u}u + \bar{d}d | N \rangle $, the value of the sigma term can be related to the baryon octet masses, as explained in Sec.~\ref{Sec2}. 
Evaluating Eq.~\eqref{Eq:II.A-3}, one finds $\sigma_{\pi N} \approx 27$~MeV, a much lower value than the dispersive determinations of that time. 
From that analysis one concludes that a value of the sigma term around $60$~MeV implied a {\it large} violation of the Zweig rule \cite{Cheng:1975wm}.
In the following years, most of the discussions about the sigma term turned around this incompatibility of a small strange quark condensate inside the nucleon and a large value of $\sigma_{\pi N}$, which was referred some times as the {\it strangeness puzzle}. 
Specially relevant for the literature of the sigma term was the article \cite{Gasser:1980sb}, where the author calculated corrections to the Eq.~\eqref{Eq:II.A-3} using Chiral Perturbation Theory and claimed the incompatibility of a small $\langle N | \bar{s}s | N \rangle$ and a $\sigma_{\pi N}\approx 60$~MeV. 

The tension between the dispersive results and the expected value assuming exact Zweig rule was reduced after the incorporation of the non-analytic corrections calculated in \cite{Pagels:1972kh}. 
They reduced the value of the sigma term obtained from the extrapolation to the Cheng-Dashen point in $\Delta_\sigma \approx 14$~MeV decreasing, therefore, the discrepancy. 
Although this correction was reported soon after the article by Cheng and Dashen, it was not included in any of the dispersive analyses mentioned before. 
It was the work of Ref.~\cite{Koch:1982pu}, where the author used hyperbolic dispersion relations, that established a dispersive result for the sigma term, $\sigma_{\pi N} \approx 50(8)$~MeV\footnote{The author provides the value of $\Sigma$. This number considers the reduction by $\Delta_\sigma$.}, that was not so far from the estimated value using $\langle N | \bar{s}s | N \rangle = 0$, $\sigma_{\pi N} \approx 35(5)$~MeV \cite{Gasser:1980sb}. 
\\

{\it The controversy on $\Delta_\sigma$}
\\

In 1988 appeared in the literature the first chiral EFT calculation with nucleons \cite{Gasser:1987rb}. 
It was applied to study $\pi N$ scattering and related low-energy theorems. 
One of the quantities studied was $\Delta_\sigma = \sigma(t = 2 M_\pi^2) - \sigma(t = 0)$. 
This calculation offered the possibility of improving the result of Ref.~\cite{Pagels:1972kh}, since the EFT formalism provides the corrections to the leading non-analytic behavior.
What the authors found is that the calculation of Pagels and Pardee was incorrect, and that the value of $\Delta_\sigma$ at leading non-analytic order was actually a factor two smaller. 
Including the complete leading one-loop corrections, they obtained $\Delta_\sigma = 4.6$~MeV, what increases again the difference between the dispersion theory calculations and the estimations using exact fulfillment of the Zweig rule.

This difference was again reduced in 1991 after the work of Ref.~\cite{Gasser:1990ap,Gasser:1990ce}, where the authors studied thoroughly all the elements involved in the determination of $\sigma_{\pi N}$. 
Starting form the Cheng-Dashen theorem, they could relate the value of the sigma term to two of the coefficients that appear in the so-called subthreshold expansion, as explained in Sec.\ref{Sec2} [see Eq. \eqref{Eq:GLS1}]. 
Using dispersion theory they could write these coefficients as functions the $\pi N$ threshold parameters. 
Taking the results for this quantities from different partial wave analyses, they extracted a value of $\sigma_{\pi N} = 45(8)$~MeV \cite{Gasser:1990ce}. 
In that determination, one of the main contributions was the update on the value of $\Delta_\sigma$ with dispersion theory \cite{Gasser:1990ap}. 
What they found is that the chiral calculation underestimated $\Delta_\sigma$ in approximately $10$~MeV, just simply because chiral EFT is not able to reproduce the dependence with energy of the strong $I=0$ $\pi \pi$ interaction in the $t$-channel of $\pi N$ at leading one-loop order.
According to their calculation, the correct value was $\Delta_\sigma = 15.2(4)$~MeV, what is remarkably close to the first (erroneous) calculation presented in Ref.~\cite{Pagels:1972kh}.
This update moved down the value of the sigma term again and made it more compatible with a small strangeness content in the nucleon. 
Still, a value of $\sigma_{\pi N} = 45$~MeV, implied a contribution of around a $20\%$ compared to the light quarks one, but it was considered acceptable at that time. 
\\

{\it Effective field theory calculations of the sigma term}
\\

After the result of Ref.~\cite{Gasser:1990ce}, there were many chiral calculations reporting a wide variety of results for the sigma term. 
Most of them were not focused on the extraction of $\sigma_{\pi N}$, and just were showing the value related to the LEC $c_1$ [see Eq.~\eqref{Eq:LEC_c1}] they were extracting in fits to $\pi N$ scattering or other sources. 
These calculations were mostly done in a heavy baryon approximation \cite{Fettes:1998ud,Fettes:2000bb} and their results ranged from negative $\sigma_{\pi N}$ to $\sigma_{\pi N} = 94$~MeV. 
Given the unacceptable values some of these extractions, sometimes they were combined with sum rules, where the scattering lengths were provided by ChEFT and the extrapolation to the Cheng-Dashen point by the corresponding sum rule \cite{Fettes:2000bb}. 
As mentioned before, Ref.~\cite{Gasser:1990ap} pointed out the difficulties of chiral EFT to calculate certain quantities related to the sigma term, as $\Delta_\sigma$, at least at leading one-loop order. 
This was supported in 2001 by Becher and Leutwyler, when they calculated the $\pi N$ scattering amplitude up to $\mathcal{O}(p^4)$ (sub-leading one-loop order calculation) in a particular Lorentz invariant formulation of ChEFT with baryons Ref.~\cite{Becher:1999he}.
There, the authors found that, the sub-leading loop diagrams bring the chiral calculation of $\Delta_\sigma$ into agreement with the dispersion theory ones.
However, the most relevant result of this work for chiral EFT refers to the impossibility of connecting the physical region of $\pi N$ scattering with the subthreshold region.
In other words, by fixing the LECs with the value of the subthreshold coefficients, the authors were not able to reproduce the scattering data. 
This was specially problematic for chiral EFT approaches, since the information in the physical region is the input used to fix the LECs and extrapolate the amplitude $\bar{D}^+$ to the Cheng-Dashen point to determine $\sigma_{\pi N}$. 
According to that work, chiral EFT could not be used to extract information in the subthreshold region and, therefore, was not suited to extract the sigma term.
\\

{\it Modern analyses and the new scenario for the sigma terms}
\\

The apparent impossibility of chiral EFT to connect the subthreshold region with the physical one was not a problem affecting only to the extraction of $\sigma_{\pi N}$, but also to the applicability of this approach to nuclear interactions.
In 2011 this issue was investigated in \cite{Alarcon:2012kn} with an $\mathcal{O}(p^3)$ calculation of $\pi N$ scattering, and it was found that this problem could be attributed to two factors. 
The first one is the incorrect analytic structure of the $\pi N$ amplitude of the previous approaches. 
The second, the necessity of including the $\Delta(1232)$ as an explicit degree of freedom to achieve the correct separation of scales demanded in an effective theory approach. 
When these limitations are solved, the chiral amplitude could reproduce the dispersive results in the subthreshold region when using the $\pi N$ phase shift to fix the LECs, as was demonstrated in \cite{Alarcon:2012kn}.
In that work, the subhtreshold coefficients $\bar{d}_{00}^+$, $\bar{d}_{01}^+$ and the difference $\Delta_D - \Delta_\sigma$ was correctly reproduced, showing that at leading-loop order chiral EFT is also able to extract $\sigma_{\pi N}$ by means of the Cheng-Dashen theorem (in the way proposed by Ref.~\cite{Gasser:1990ce}). 
The approach presented in \cite{Alarcon:2012kn} was used to calculate the sigma term and to study the phenomenology of $\pi N$ scattering, and the extracted value was $\sigma_{\pi N} = 59(7)$~MeV \cite{Alarcon:2011zs}.  
In this extraction the scattering lengths obtained in modern $\pi$-atoms spectroscopy experiments where used to check the reliability of the phase shifts used in the fits.
It was shown that the updated result for the scalar-isoscalar $\pi N$ scattering length $a_{0+}^+$ clearly favored the input that leads to $\sigma_{\pi N} =  59(7)$~MeV. 
This was similar to the conclusions of Ref.~\cite{Pavan:2001wz}, where the authors claimed that modern $\pi N$ scattering favoured a sigma term $\sigma_{\pi N} \approx 60$~MeV.

Although modern deteminations of the $\pi N$ scattering lengths pointed to a value of $\sigma_{\pi N} \sim 60$~MeV, the problem of a large strange quark content in the nucleon was still present. 
This issue was investigated in \cite{Alarcon:2012nr} just after the publication of \cite{Alarcon:2011zs}.
There, it was shown that the relation of the baryon masses with the matrix element $\langle N |\hat{m}( \bar{u}u + \bar{d}d -2 \bar{s}s )|N \rangle$ receives sizeable higher order corrections, so that the leading order $SU(3)$ breaking contribution used in the past was not accurate enough. 
Once these the next-to-leading order was included using $SU(3)$ chiral EFT \cite{Alarcon:2012nr}, it was shown that a sigma term of the order of $60$~MeV was perfectly compatible with a negligible strangeness in the nucleon (small $\sigma_s$). 
With the solution of the strangeness puzzle, emerged a new scenario for the sigma terms, where the modern experimental information favored a value of $\sigma_{\pi N} \approx 60$~MeV with a negligible strangeness content in the nucleon (in contrast with the previous situation of a strangeness of $\approx 20\%$).

The thesis of a large value of the sigma term driven by the modern extraction of the $\pi N$ scattering lengths was reinforced after the publication of Ref.~\cite{Hoferichter:2015dsa}.
In that work, using Roy-Steiner equations, and including the leading-order isospin breaking effects in chiral EFT, the authors determined a value of the sigma term of $\sigma_{\pi N} = 59.1(3.5)$~MeV. 
In this determination, the recent extraction of the $\pi N$ scattering lengths was also necessary to distinguish between a large and a low value of the sigma term.
The excellent agreement of this result with Ref.~\cite{Alarcon:2011zs} gave a strong support to the claims of a large value of $\sigma_{\pi N}$, given the noticeable differences between the chiral and dispersive approaches\footnote{Notice that the approach of \cite{Hoferichter:2015dsa} is much closer to \cite{Gasser:1990ce} than to \cite{Alarcon:2011zs}. }.
\\

{\it Lattice QCD determinations at the physical point and re-analysis of the strangeness puzzle}
\\

With the publication of \cite{Hoferichter:2015dsa} it seemed that the sigma term puzzle was finally solved: Different approaches that used modern data converged in values around $\sigma_{\pi N} \approx 60$~MeV, and this large value was not in conflict with a small strangeness content in the nucleon anymore. 
However, soon after the publication of \cite{Hoferichter:2015dsa}, several lattice groups started to get results for the sigma term at the physical point for fist time in the literature. 
In these works both methods, the Hellmann-Feynman theorem \cite{Durr:2015dna} and the direct computation of the three-point functions \cite{Yang:2015uis,Abdel-Rehim:2016won,Bali:2016lvx}, were used to calculate the matrix element $ \langle N |\hat{m}( \bar{u}u + \bar{d}d )|N \rangle$.
The results of these analyses turned out to agree in a small value of the sigma term: $38(3)(3)$~MeV \cite{Durr:2015dna}, $44.4(3.2)(4.5)$~MeV \cite{Yang:2015uis}, $37.22(2.57)\left(^{+0.99}_{-0.63}\right)$~MeV \cite{Abdel-Rehim:2016won} and $35(6)$~MeV \cite{Bali:2016lvx}, supporting the value reported in Ref.~\cite{Gasser:1990ce}.

In addition, the solution of the strangeness puzzle was questioned in Ref.~\cite{Leutwyler:2015jga} in connection with the Gell-Mann-Okubo (GMO) formula. 
The observation lies in fact that the GMO mass formula, based on a leading-order calculation in the symmetry breaking parameter $m_s - \hat{m}$, works with high accuracy for the baryon octet. 
Therefore, one can wonder how can fail that much in the case of $\sigma_0$, Eq.~\eqref{Eq:sigma0}.
This question was answered a few years later in \cite{Fernando:2018jrz}, showing that the corrections to $\Delta_{GMO}$ and $\sigma_0$ are related by chiral symmetry. 
A relatively large correction to $\sigma_0$ is found once the next-to-leading order correction to the mass formula is included (in the latter case, improving its accuracy). 
There was, however, an important difference between both corrections. 
While the scaling of the correction to the GMO mass formula is $1/N_c$, the correction to $\sigma_0$ scales as $N_c$.
\\

{\it Present status of the sigma term puzzle and possible solutions}
\\

The fact that lattice obtained a value of the sigma term much lower than the modern phenomenological extractions re-opened the apparently overcome debate between a low and a large value of $\sigma_{\pi N}$. 
There was, however, an important difference between the current situation and the arguments in the past. 
In the past, the discrepancy on the value of the sigma term showed up when comparing different phenomenological extractions.
Nowadays, the different phenomenological extractions of the sigma term seem to converge in a large value, $\sigma_{\pi N} \approx 60$~MeV, and the disagreement appear now between these and the lattice calculations. 
  
Subsequent phenomenological and lattice analyses reinforced the current discrepancy. 
On the phenomenological side, for example, a fit to $\pi N$ scattering data using Roy-Steiner equations \cite{RuizdeElvira:2017stg} extracted a value of the sigma term of $\sigma_{\pi N} = 58(5)$~MeV, in agreement with previous phenomenological analyses. 
Later, there was another independent extraction based on pionic atoms data, where a value of $\sigma_{\pi N} = 57(7)$~MeV was reported \cite{Friedman:2019zhc}, this time based on the value of the $\pi N$ isovector scattering length.
On the lattice side, calculations like the one in Ref.~\cite{Alexandrou:2017qyt,Borsanyi:2020bpd}, gave support to a value around $40$~MeV, although that low value is at odds with the scalar-isoscalar scattering length reported by Ref.~\cite{Baru:2011bw}.

At this point, and given the consistency between modern phenomenological analyses on one side, and the different lattice calculations on the other, it seems one needs to reexamine the common sources that led to these values. 
From the phenomenological side, one has to consider that most of the modern determinations of $\sigma_{\pi N}$ rely on $\pi$-atom spectroscopy, in particular on the scattering lengths.
The latter are determined from the energy shift and width of the $1s$ level of $\pi H$.
The connection of these quantities to the isoscalar and isovector scattering lengths is based on a chiral EFT calculation, which dominates the systematic uncertainty of the extraction. 
I think it is worth to review these calculations at some point, specially what refers to $\pi N N$ scattering, the inclusion of isospin violation and Coulomb corrections.
Another possible source of error could be the $\pi N$ data base. 
Some authors claim that the $\pi N$ database is inconsistent \cite{Matsinos:2016fcd}, and it would be relevant for the sigma term extraction to investigate this issue. 
In what regards to the lattice calculations, a reexamination of the extractions at the physical point would be also desirable, although specific suggestions would correspond to the experts in that field. 
In any case, despite its negative aspects, the sigma term puzzle will force both communities to improve their state-of-the-art calculations, and
this will be mutually beneficial to both parties, as happened in the past with the phenomenological extractions.

 \begin{figure}[H]
 \begin{center}
 \includegraphics[width=.7\textwidth]{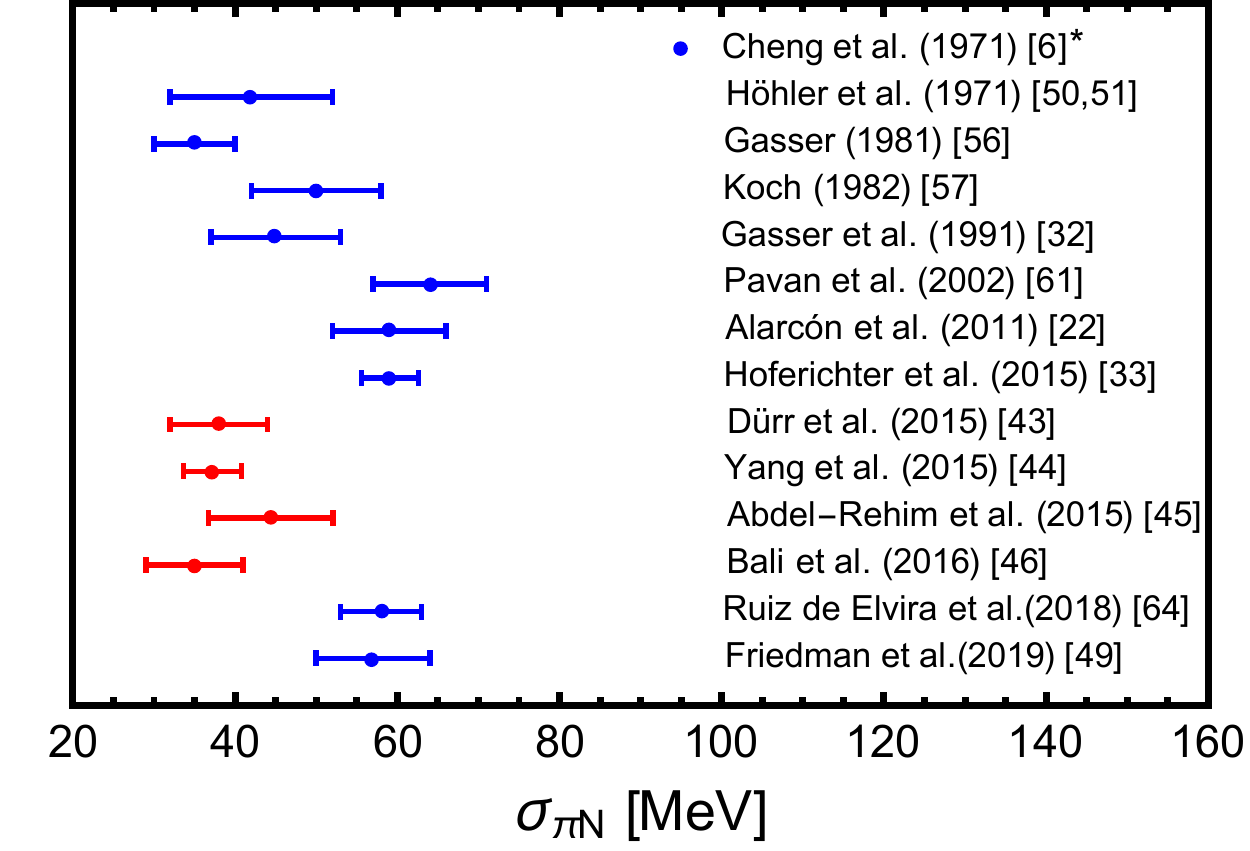} 
 \caption{Summary of some of the results mentioned in Sec.~\ref{Sec:sigma-term_puzzle}. Blue dots correspond to phenomenological determinations ($SU(3)$ breaking models, dispersion theory, chiral EFT or pionic atoms). Red dots correspond to the lattice QCD calculations at the physical point.$^*$ This number is corrected by the actual value of $\Delta_\sigma$, not considered in the original calculation by Cheng and Dashen. }
 \label{Fig:Summary_sigma_terms}
 \end{center}
\end{figure} 

\section{Summary and Conclusions}

The pion-nucleon sigma term is an important quantity in many areas of particle and nuclear physics. 
It is related to the chiral symmetry breaking of the strong interactions, the origin of the mass of the ordinary matter and searches of physics beyond the standard model. 
Low-energy theorems relate $\sigma_{\pi N}$ to the $\pi N$ scattering amplitude, what made it accesible though experimental data.
Although much effort has been invested in its determination since then, nowadays its size is still not clear.
Starting from $\sigma_{\pi N} \sim 30$~MeV, its accepted value has been oscillating from small to large during the last fifty years. 
At the beginning, it was attributed to the quality of the experimental data used in dispersive approaches or convergence problems of other approaches. 
With the time, the determinations from different methods became more robust and, including modern experimental information (being the $\pi N$ scalar scattering lengths the most relevant one), the later phenomenological extractions converged to a value close to $60$~MeV.
This value, that in the past was related to a huge strangeness content in the nucleon, has been proven to be consistent with a small $\sigma_s$.
The convergence around $60$~MeV by so different methods as dispersion theory, chiral effective field theory or pionic atoms, that use different theoretical approaches and different input, strengthened the modern phenomenological determinations.
However, recent numerical calculations in lattice QCD at the physical point challenged this picture provided from the phenomenological side.
Various lattice groups agree on a value of $\sigma_{\pi N} \sim 40$~MeV using different methods, setting a discrepancy between lattice and phenomenological extractions that continues nowadays.
This reminds the scenario of the early days of the sigma term, with the important difference that now all modern phenomenological extractions agree. The discrepancy shows up only when comparing phenomenological and lattice calculations. 
The origin of this difference is so far unknown, but the puzzle it raises becomes a great opportunity to investigate and improve the approaches involved in the determination of this important quantity.

\section{Acknowledgements}

I would like to thank C. Alexandrou and N. Kaiser for allowing me to use the figures \ref{Fig:Lattice} and \ref{Fig:chiral_condensate}. I want to thank also J. A. Oller for a careful reading of the manuscript.

%\section{\refname}


\begin{thebibliography}{99}

\bibitem{Adler}
S.~L.~Adler,
%``Consistency conditions on the strong interactions implied by a partially conserved axial vector current,''
Phys. Rev. \textbf{137}, B1022-B1033 (1965); %``Consistency conditions on the strong interactions implied by a partially conserved axial-vector current. II,''
Phys. Rev. \textbf{139}, B1638-B1643 (1965).


\bibitem{Weinberg}
S.~Weinberg,
%``Pion scattering lengths,''
Phys. Rev. Lett. \textbf{17}, 616-621 (1966).

\bibitem{GellMann:1968rz}
M.~Gell-Mann, R.~J.~Oakes and B.~Renner,
%``Behavior of current divergences under SU(3) x SU(3),''
Phys. Rev. \textbf{175} (1968), 2195-2199.


\bibitem{Reya:1974gk}
E.~Reya,
%``Chiral symmetry breaking and meson - nucleon sigma commutators: A Review,''
Rev. Mod. Phys. \textbf{46} (1974), 545-580.

\bibitem{Fubini&Furlan}
S.~Fubini and G.~Furlan,
%``Dispersion Theory of Low-energy Limits''
Ann. of Phys. \textbf{48} (1968), 322-365.

\bibitem{Cheng:1970mx}
T.~P.~Cheng and R.~F.~Dashen,
%``Is SU(2) x SU(2) a better symmetry than SU(3)?,''
Phys. Rev. Lett. \textbf{26} (1971), 594.

\bibitem{Brown:1971pn}
L.~S.~Brown, W.~J.~Pardee and R.~D.~Peccei,
%``Adler-Weisberger theorem reexamined,''
Phys. Rev. D \textbf{4} (1971), 2801-2810.



\bibitem{Bernard:1996nu}
V.~Bernard, N.~Kaiser and U.~G.~Meissner,
%``On the analysis of the pion - nucleon sigma term: The Size of the remainder at the Cheng-Dashen point,''
Phys. Lett. B \textbf{389} (1996), 144-148.


\bibitem{Pagels:1972kh}
H.~Pagels and W.~J.~Pardee,
%``Nonanalytic behavior of the sigma term in pi-n scattering,''
Phys. Rev. D \textbf{4} (1971), 3335-3337.




\bibitem{Crewther:1972kn}
R.~J.~Crewther,
%``Nonperturbative evaluation of the anomalies in low-energy theorems,''
Phys. Rev. Lett. \textbf{28}, 1421 (1972).


\bibitem{Chanowitz:1972vd}
M.~S.~Chanowitz and J.~R.~Ellis,
%``Canonical Anomalies and Broken Scale Invariance,''
Phys. Lett. B \textbf{40}, 397-400 (1972).


\bibitem{Collins:1976yq}
J.~C.~Collins, A.~Duncan and S.~D.~Joglekar,
%``Trace and Dilatation Anomalies in Gauge Theories,''
Phys. Rev. D \textbf{16}, 438-449 (1977).

\bibitem{Shifman:1978zn}
M.~A.~Shifman, A.~I.~Vainshtein and V.~I.~Zakharov,
%``Remarks on Higgs Boson Interactions with Nucleons,''
Phys. Lett. B \textbf{78}, 443-446 (1978).




\bibitem{Hill:2011be}
R.~J.~Hill and M.~P.~Solon,
%``Universal behavior in the scattering of heavy, weakly interacting dark matter on nuclear targets,''
Phys. Lett. B \textbf{707} (2012), 539-545.

\bibitem{Crivellin:2013ipa}
A.~Crivellin, M.~Hoferichter and M.~Procura,
%``Accurate evaluation of hadronic uncertainties in spin-independent WIMP-nucleon scattering: Disentangling two- and three-flavor effects,''
Phys. Rev. D \textbf{89}, 054021 (2014).

\bibitem{Bottino}
A.~Bottino, F.~Donato, N.~Fornengo and S.~Scopel,
%``Size of the neutralino nucleon cross-section in the light of a new determination of the pion nucleon sigma term,''
Astropart. Phys. \textbf{18}, 205-211 (2002); %``Interpreting the recent results on direct search for dark matter particles in terms of relic neutralino,''
Phys. Rev. D \textbf{78}, 083520 (2008).



\bibitem{Ellis:2008hf}
J.~R.~Ellis, K.~A.~Olive and C.~Savage,
%``Hadronic Uncertainties in the Elastic Scattering of Supersymmetric Dark Matter,''
Phys. Rev. D \textbf{77}, 065026 (2008).



\bibitem{Elhatisari:2016owd}
S.~Elhatisari, N.~Li, A.~Rokash, J.~M.~Alarc\'on, D.~Du, N.~Klein, B.~n.~Lu, U.~G.~Mei\ss{}ner, E.~Epelbaum and H.~Krebs, \textit{et al.}
%``Nuclear binding near a quantum phase transition,''
Phys. Rev. Lett. \textbf{117}, no.13, 132501 (2016).


\bibitem{Berengut:2013nh}
J.~C.~Berengut, E.~Epelbaum, V.~V.~Flambaum, C.~Hanhart, U.~G.~Meissner, J.~Nebreda and J.~R.~Pelaez,
%``Varying the light quark mass: impact on the nuclear force and Big Bang nucleosynthesis,''
Phys. Rev. D \textbf{87}, no.8, 085018 (2013).

\bibitem{Weise:2012yv}
W.~Weise,
%``Nuclear chiral dynamics and phases of QCD,''
Prog. Part. Nucl. Phys. \textbf{67}, 299-311 (2012).


\bibitem{Kaiser:2007nv}
N.~Kaiser, P.~de Homont and W.~Weise,
%``In-medium chiral condensate beyond linear density approximation,''
Phys. Rev. C \textbf{77}, 025204 (2008).

\bibitem{Oller:2001sn}
J.~A.~Oller,
%``Chiral Lagrangians at finite density,''
Phys. Rev. C \textbf{65}, 025204 (2002).

\bibitem{Meissner:2001gz}
U.~G.~Meissner, J.~A.~Oller and A.~Wirzba,
%``In-medium chiral perturbation theory beyond the mean field approximation,''
Annals Phys. \textbf{297}, 27-66 (2002).

\bibitem{Oller:2009zt}
J.~A.~Oller, A.~Lacour and U.~G.~Meissner,
%``Chiral Effective Field Theory for Nuclear Matter with long- and short-range Multi-Nucleon Interactions,''
J. Phys. G \textbf{37}, 015106 (2010).

\bibitem{Lacour:2010ci}
A.~Lacour, J.~A.~Oller and U.~G.~Meissner,
%``The Chiral quark condensate and pion decay constant in nuclear matter at next-to-leading order,''
J. Phys. G \textbf{37}, 125002 (2010).

\bibitem{Lacour:2009ej}
A.~Lacour, J.~A.~Oller and U.~G.~Meissner,
%``Non-perturbative methods for a chiral effective field theory of finite density nuclear systems,''
Annals Phys. \textbf{326}, 241-306 (2011).

\bibitem{Oller:2019ssq}
J.~A.~Oller,
%``An in-medium chiral power-counting scheme for nuclear matter and some applications,''
J. Phys. G \textbf{46}, no.7, 073001 (2019).




\bibitem{Alarcon:2011zs}
J.~M.~Alarcon, J.~Martin Camalich and J.~A.~Oller,
%``The chiral representation of the $\pi N$ scattering amplitude and the pion-nucleon sigma term,''
Phys. Rev. D \textbf{85}, 051503 (2012).


\bibitem{Alarcon:2012nr}
J.~M.~Alarcon, L.~S.~Geng, J.~Martin Camalich and J.~A.~Oller,
%``The strangeness content of the nucleon from effective field theory and phenomenology,''
Phys. Lett. B \textbf{730}, 342-346 (2014).


\bibitem{Fernando:2018jrz}
I.~P.~Fernando, J.~M.~Alarcon and J.~L.~Goity,
%``Baryon masses and $\sigma$ terms in SU(3) BChPT x 1/N$_c$,''
Phys. Lett. B \textbf{781}, 719-722 (2018).


\bibitem{Cheng:1975wm}
T.~P.~Cheng,
%``The Zweig Rule and the pi n Sigma Term,''
Phys. Rev. D \textbf{13}, 2161 (1976).

\bibitem{Neeman:1961jhl}
Y.~Ne'eman,
%``Derivation of strong interactions from a gauge invariance,''
Nucl. Phys. \textbf{26}, 222-229 (1961).

\bibitem{GellMann:1962xb}
M.~Gell-Mann,
%``Symmetries of baryons and mesons,''
Phys. Rev. \textbf{125}, 1067-1084 (1962).


\bibitem{GellMann:1964tf}
M.~Gell-Mann,
%``The Symmetry group of vector and axial vector currents,''
Physics Physique Fizika \textbf{1}, 63-75 (1964).


\bibitem{VonHippel:1970uz}
F.~Von Hippel and J.~K.~Kim,
%``Nature of su(3) x su(3) symmetry breaking - results from a systematic test of the soft-meson theorems,''
Phys. Rev. D \textbf{1}, 151-164 (1970)
[erratum: Phys. Rev. D \textbf{3}, 2923-2923 (1971)].


\bibitem{Gasser:1988jt}
J.~Gasser, H.~Leutwyler, M.~P.~Locher and M.~E.~Sainio,
%``Extracting the Pion - Nucleon $\Sigma$ Term From Data,''
Phys. Lett. B \textbf{213} (1988), 85-90.



\bibitem{Gasser:1990ap}
J.~Gasser, H.~Leutwyler and M.~E.~Sainio,
%``Form-factor of the sigma term,''
Phys. Lett. B \textbf{253}, 260-264 (1991).

\bibitem{Oller:2007xd}
J.~A.~Oller and L.~Roca,
%``Scalar radius of the pion and zeros in the form factor,''
Phys. Lett. B \textbf{651}, 139-146 (2007).

\bibitem{Donoghue:1990xh}
J.~F.~Donoghue, J.~Gasser and H.~Leutwyler,
%``The Decay of a Light Higgs Boson,''
Nucl. Phys. B \textbf{343}, 341-368 (1990).


\bibitem{Gasser:1990ce}
J.~Gasser, H.~Leutwyler and M.~E.~Sainio,
%``Sigma term update,''
Phys. Lett. B \textbf{253}, 252-259 (1991).


\bibitem{Hoferichter:2015dsa}
M.~Hoferichter, J.~Ruiz de Elvira, B.~Kubis and U.~G.~Mei\ss{}ner,
%``High-Precision Determination of the Pion-Nucleon \ensuremath{\sigma} Term from Roy-Steiner Equations,''
Phys. Rev. Lett. \textbf{115}, 092301 (2015).

\bibitem{Becher:1999he}
T.~Becher and H.~Leutwyler,
%``Baryon chiral perturbation theory in manifestly Lorentz invariant form,''
Eur. Phys. J. C \textbf{9}, 643-671 (1999).





\bibitem{Altarelli:1971kh}
G.~Altarelli, N.~Cabibbo and L.~Maiani,
%``Magnitude of the sigma-term, chiral symmetry and scale invariance,''
Phys. Lett. B \textbf{35} (1971), 415-418.

\bibitem{Olsson:1979ee}
M.~G.~Olsson and E.~T.~Osypowski,
%``A Rigorous Derivation of the Altarelli-cabibbo-maiani Relation,''
J. Phys. G \textbf{6} (1980), 423.

\bibitem{Olsson:1999jt}
M.~G.~Olsson,
%``The Nucleon sigma term from threshold parameters,''
Phys. Lett. B \textbf{482}, 50-56 (2000).

\bibitem{Gotta:2008zza}
D.~Gotta, F.~Amaro, D.~F.~Anagnostopoulos, S.~Biri, D.~S.~Covita, H.~Gorke, A.~Gruber, M.~Hennebach, A.~Hirtl and T.~Ishiwatari, \textit{et al.}
%``Pionic hydrogen,''
Lect. Notes Phys. \textbf{745}, 165-186 (2008).


\bibitem{Baru:2011bw}
V.~Baru, C.~Hanhart, M.~Hoferichter, B.~Kubis, A.~Nogga and D.~R.~Phillips,
%``Precision calculation of threshold $\pi^-d$ scattering, $\pi$N scattering lengths, and the GMO sum rule,''
Nucl. Phys. A \textbf{872}, 69-116 (2011).


\bibitem{Gasser&Leutwyler}
J.~Gasser and H.~Leutwyler,
%``Low-Energy Theorems as Precision Tests of QCD,''
Phys. Lett. B \textbf{125}, 325-329 (1983);%``Chiral Perturbation Theory to One Loop,''
Annals Phys. \textbf{158}, 142 (1984);%``Chiral Perturbation Theory: Expansions in the Mass of the Strange Quark,''
Nucl. Phys. B \textbf{250}, 465-516 (1985).


\bibitem{Alarcon:2012kn}
J.~M.~Alarcon, J.~Martin Camalich and J.~A.~Oller,
%``Improved description of the $\pi N$-scattering phenomenology in covariant baryon chiral perturbation theory,''
Annals Phys. \textbf{336}, 413-461 (2013).



\bibitem{Hellmann-Feynman}
P. G\"uttinger, ``Das verhalten von atomen im magnetischen drehfeld", Zeitschrift f\"ur Physik 73, 169-184 (1932); W. Pauli, {\it Principles of Wave Mechanics}, Vol. 24 (Springer, 1933) p. 162; H. Hellmann, Einf\"uhrung in die Quantenchemie (Franz
Deuticke, 1937) p. 285, OL21481721M; R. P. Feynman, ``Forces in molecules", Phys. Rev. 56,
340-343 (1939).

\bibitem{Abdel-Rehim:2016won}
A.~Abdel-Rehim \textit{et al.} [ETM],
%``Direct Evaluation of the Quark Content of Nucleons from Lattice QCD at the Physical Point,''
Phys. Rev. Lett. \textbf{116}, no.25, 252001 (2016).


\bibitem{Durr:2015dna}
S.~Durr, Z.~Fodor, C.~Hoelbling, S.~D.~Katz, S.~Krieg, L.~Lellouch, T.~Lippert, T.~Metivet, A.~Portelli and K.~K.~Szabo, \textit{et al.}
%``Lattice computation of the nucleon scalar quark contents at the physical point,''
Phys. Rev. Lett. \textbf{116}, no.17, 172001 (2016).

\bibitem{Yang:2015uis}
Y.~B.~Yang \textit{et al.} [xQCD],
%``$\pi$N and strangeness sigma terms at the physical point with chiral fermions,''
Phys. Rev. D \textbf{94}, no.5, 054503 (2016).

\bibitem{Bali:2016lvx}
G.~S.~Bali \textit{et al.} [RQCD],
%``Direct determinations of the nucleon and pion $\sigma$ terms at nearly physical quark masses,''
Phys. Rev. D \textbf{93}, no.9, 094504 (2016).

\bibitem{Alexandrou:2017qyt}
C.~Alexandrou, M.~Constantinou, P.~Dimopoulos, R.~Frezzotti, K.~Hadjiyiannakou, K.~Jansen, C.~Kallidonis, B.~Kostrzewa, G.~Koutsou and M.~Mangin-Brinet, \textit{et al.}
%``Nucleon scalar and tensor charges using lattice QCD simulations at the physical value of the pion mass,''
Phys. Rev. D \textbf{95}, no.11, 114514 (2017)
[erratum: Phys. Rev. D \textbf{96}, no.9, 099906 (2017)].

\bibitem{Borsanyi:2020bpd}
S.~Borsanyi, Z.~Fodor, C.~Hoelbling, L.~Lellouch, K.~K.~Szabo, C.~Torrero and L.~Varnhorst,
%``Ab-initio calculation of the proton and the neutron's scalar couplings for new physics searches,''
[arXiv:2007.03319 [hep-lat]].


\bibitem{Friedman:2019zhc}
E.~Friedman and A.~Gal,
%``The pion-nucleon \ensuremath{\sigma} term from pionic atoms,''
Phys. Lett. B \textbf{792}, 340-344 (2019).



\bibitem{Hoehler:1971km}
G.~Hoehler, H.~P.~Jakob and R.~Strauss,
%``The magnitude of the sigma-commutator term in current algebra,''
Phys. Lett. B \textbf{35}, 445-449 (1971).

\bibitem{Hoehler:1972gb}
G.~Hoehler, H.~P.~Jakob and R.~Strauss,
%``A critical test of models for the low energy pi n amplitude,''
Nucl. Phys. B \textbf{39}, 237-266 (1972).





\bibitem{Nielsen:1974qz}
H.~Nielsen and G.~C.~Oades,
%``Low-energy pi n partial waves, expansions of the pi n invariant amplitudes about nu=0, t=0 and the value of the current algebra sigma term,''
Nucl. Phys. B \textbf{72}, 310-320 (1974).

\bibitem{Hite:1974wz}
G.~E.~Hite and R.~J.~Jacob,
%``An Interior Dispersion Relation Determination of the pi n Sigma-Term,''
Phys. Lett. B \textbf{53}, 200-202 (1974).

\bibitem{Langbein:1974rz}
W.~Langbein,
%``Analysis of the pi n Invariant Amplitudes in Terms of Algebraic Functions and New Determination of the pi n Low-Energy Parameters,''
Nucl. Phys. B \textbf{94}, 519-547 (1975).

\bibitem{Chao:1975sk}
Y.~A.~Chao, R.~E.~Cutkosky, R.~L.~Kelly and J.~W.~Alcock,
%``Evaluation of the Pion-Nucleon Sigma Term,''
Phys. Lett. B \textbf{57}, 150-154 (1975).



\bibitem{Gasser:1980sb}
J.~Gasser,
%``Hadron Masses and Sigma Commutator in the Light of Chiral Perturbation Theory,''
Annals Phys. \textbf{136}, 62 (1981).


\bibitem{Koch:1982pu}
R.~Koch,
%``A New Determination of the pi N Sigma Term Using Hyperbolic Dispersion Relations in the (nu**2, t) Plane,''
Z. Phys. C \textbf{15}, 161-168 (1982).


\bibitem{Gasser:1987rb}
J.~Gasser, M.~E.~Sainio and A.~Svarc,
%``Nucleons with Chiral Loops,''
Nucl. Phys. B \textbf{307}, 779-853 (1988).


\bibitem{Fettes:1998ud}
N.~Fettes, U.~G.~Meissner and S.~Steininger,
%``Pion - nucleon scattering in chiral perturbation theory. 1. Isospin symmetric case,''
Nucl. Phys. A \textbf{640}, 199-234 (1998).

\bibitem{Fettes:2000bb}
N.~Fettes and U.~G.~Meissner,
%``Pion - nucleon scattering in an effective chiral field theory with explicit spin 3/2 fields,''
Nucl. Phys. A \textbf{679}, 629-670 (2001).


\bibitem{Pavan:2001wz}
M.~M.~Pavan, I.~I.~Strakovsky, R.~L.~Workman and R.~A.~Arndt,
%``The Pion nucleon Sigma term is definitely large: Results from a G.W.U. analysis of pi nucleon scattering data,''
PiN Newslett. \textbf{16}, 110-115 (2002).


\bibitem{Leutwyler:2015jga}
H.~Leutwyler,
%``Theoretical aspects of Chiral Dynamics,''
PoS \textbf{CD15}, 022 (2015).


\bibitem{RuizdeElvira:2017stg}
J.~Ruiz de Elvira, M.~Hoferichter, B.~Kubis and U.~G.~Mei\ss{}ner,
%``Extracting the $\sigma$-term from low-energy pion-nucleon scattering,''
J. Phys. G \textbf{45}, no.2, 024001 (2018).


\bibitem{Matsinos:2016fcd}
E.~Matsinos and G.~Rasche,
%``Systematic effects in the low-energy behavior of the current SAID solution for the pion-nucleon system,''
Int. J. Mod. Phys. E \textbf{26}, no.03, 1750002 (2017).
  
  
 \end{thebibliography}
\end{document}